\documentclass[a4paper,12pt]{article}

\usepackage{epsf,amssymb,amsmath,dsfont,bbold}
\usepackage{epsfig}
\usepackage{hyperref}
\usepackage{multirow}
\usepackage{color}
\usepackage{cite}

\makeatletter
\renewcommand\section{\@startsection {section}{1}{\z@}%
                                   {-3.5ex \@plus -1ex \@minus -.2ex}
                                   {2.3ex \@plus.2ex}%
                                   {\normalfont\large\bfseries}}

\renewcommand\subsection{\@startsection{subsection}{2}{\z@}%
                                     {-3.25ex\@plus -1ex \@minus -.2ex}%
                                     {1.5ex \@plus .2ex}%
                                     {\normalfont\bfseries}}

\makeatother

\newcommand{\bea}{\begin{eqnarray}}
\newcommand{\eea}{\end{eqnarray}}
\newcommand{\be}{\begin{equation}}
\newcommand{\ee}{\end{equation}}
\newcommand{\bem}{\begin{pmatrix}}
\newcommand{\eem}{\end{pmatrix}}
\newcommand{\bl}{\begin{align}}
\newcommand{\el}{\end{align}}

\newcommand{\bi}{\begin{itemize}}
\newcommand{\ei}{\end{itemize}}

\def\p{\partial}

\def\cD{{\cal D}}

\def\cL{{\cal L}}
\def\cP{{\cal P}}
\def\cR{{\cal R}}
\def\cZ{{\cal Z}}

\newcommand{\gh}{\hat{g}}
\newcommand{\gb}{\bar{g}}
\newcommand{\wtw}{\widetilde{w}}

\newcommand{\sib}{\bar{\sigma}}
\newcommand{\sih}{\hat{\sigma}}

\newcommand{\Nb}{\bar{N}}
\newcommand{\Nh}{\hat{N}}

\newcommand{\Db}{\bar{D}}
\newcommand{\Kb}{\bar{K}}
\newcommand{\Rb}{\bar{R}}

\begin{document}
\begin{titlepage}
\begin{center}
${}$
\thispagestyle{empty}
\renewcommand{\thefootnote}{\fnsymbol{footnote}}
\vskip 1.2cm {\LARGE {\bf Impact of topology in foliated \\[1.2ex] Quantum Einstein Gravity}}
\vskip 1.2cm  { W.B.\ Houthoff${}^{1,}$\footnote{whouthoff@science.ru.nl}, A.\ Kurov${}^{1,2,}$\footnote{kurov.aleksandr@physics.msu.ru}, and F.\ Saueressig${}^1$\footnote{F.Saueressig@science.ru.nl} }\\
{\vskip 0.5cm  }
${}^1$ Institute for Mathematics, Astrophysics and Particle Physics (IMAPP) \\
Radboud University Nijmegen, 6525 AJ Nijmegen, The Netherlands \\[2ex]
${}^2$ Department of Theoretical Physics, Moscow State University,
Russia \\[2ex]
\vspace{1cm}

\begin{abstract}
\baselineskip=16pt
\noindent
We use a functional renormalization group equation tailored to the Arnowitt-Deser-Misner formulation of gravity to study the scale-dependence of Newton's coupling and the cosmological constant on a background spacetime with topology $S^1 \times S^d$. The resulting beta functions possess a non-trivial renormalization group fixed point, which may provide the high-energy completion of the theory through the asymptotic safety mechanism. The fixed point is
robust with respect to changing the parametrization of the metric fluctuations and regulator scheme. The phase diagrams show that this fixed point is connected to a classical regime through a crossover. In addition the flow may exhibit a regime of ``gravitational instability'', modifying the theory in the deep infrared. Our work complements earlier studies of the gravitational renormalization group flow on a background topology $S^1 \times T^d$ \cite{Biemans:2016rvp,Biemans:2017zca} and establishes that the flow is essentially independent of the background topology.
\end{abstract}

\end{center}
\newpage
\end{titlepage}
\tableofcontents

\def\spc{\hspace{.5pt}}

\date{\today}
\section{Introduction and motivation}
\setcounter{footnote}{0}
Asymptotic Safety, first suggested by Weinberg \cite{Weinberg:1980gg}, constitutes a mechanism for constructing a consistent and predictive quantum theory for gravity within the framework of quantum field theory. A central goal of the program is to give meaning to the path integral over (Euclidean) metrics
\be\label{partfct}
\cZ = \int \cD \gh \, \exp(-S[\gh])
\ee
with $S[\gh]$ a suitable, diffeomorphism invariant action functional. This task can be addressed along various ways \cite{Oritibook}, e.g.\ by applying continuum renormalization group methods or discrete Monte Carlo techniques.

Within the Causal Dynamical Triangulations (CDT) program, reviewed in \cite{Ambjorn:2012jv}, the action entering the partition function \eqref{partfct} is taken as the Einstein-Hilbert action and the partition sum is taken on background topologies $S^1 \times S^3$ \cite{Ambjorn:2007jv,Ambjorn:2008wc,Ambjorn:2010fv} or on topologies of the form $S^1 \times T^3$ \cite{Ambjorn:2016fbd}. $\cZ$ is then evaluated on piecewise linear geometries constructed from elementary simplices.\footnote{For related work in the context of Euclidean Dynamical Triangulations see \cite{Smit:2013wua,Ambjorn:2013eha,Rindlisbacher:2015ewa,Laiho:2016nlp}.}  The simplices provide a lattice regularization, making the partition sum finite. Removing the regulator by taking the continuum limit then requires a second order phase transition where the correlation length diverges. For CDT a candidate for such a phase transition has been identified in \cite{Ambjorn:2011cg,Ambjorn:2016mnn}, also see \cite{Ambjorn:2012ij,Ambjorn:2014gsa,Ambjorn:2016cpa} for related investigations. Moreover, random walks on CDT spacetimes exhibit manifold-like behavior for long diffusion time, indicating the presence of a classical phase \cite{Ambjorn:2004qm,Ambjorn:2005db,Ambjorn:2005qt}. These features are typically attributed to the presence of a causal structure associated with the $S^1$-factor in the topology which allows building up spacetime as a stack of spatial slices.

A second route towards asymptotic safety, reviewed in \cite{Niedermaier:2006wt,Codello:2008vh,Litim:2011cp,Percacci:2011fr,Reuter:2012id,Nagy:2012ef}, converts the partition sum \eqref{partfct} into a functional renormalization group equation (FRGE) for the effective average action $\Gamma_k$ \cite{Wetterich:1992yh,Morris:1993qb,Reuter:1993kw}. Starting from the pioneering work \cite{Reuter:1996cp}, this program has
made significant progress in demonstrating that the asymptotic safety mechanism may lead to a viable quantum theory of gravity. In particular,
the existence of a non-Gaussian fixed point (NGFP), which constitutes the key element in this program, has been demonstrated in a wide range of approximations \cite{Souma:1999at,Lauscher:2001ya,Reuter:2001ag,Litim:2003vp,Donkin:2012ud,Nagy:2013hka,Gies:2015tca,Lauscher:2002sq,Codello:2007bd,Machado:2007ea,Rechenberger:2012pm,Falls:2014tra,Falls:2016msz,Benedetti:2009gn,Gies:2016con}.  Starting from
\cite{Manrique:2009uh,Manrique:2010mq,Manrique:2010am}
renormalization group flows which resolve the difference between the background and fluctuation fields have been constructed, e.g., in \cite{Christiansen:2012rx,Codello:2013wxa,Codello:2013fpa,Christiansen:2014raa,Becker:2014jua,Becker:2014pea,Christiansen:2015rva,Labus:2016lkh,Morris:2016spn,Percacci:2016arh,Denz:2016qks} while the role of the path-integral measure has recently been discussed in \cite{Demmel:2014hla,Ohta:2016npm,Ohta:2016jvw,Falls:2017cze}.\footnote{For selected works on Asymptotic Safety in gravity-matter models see \cite{Zanusso:2009bs,Shaposhnikov:2009pv,Harst:2011zx,Dona:2013qba,Oda:2015sma,Meibohm:2015twa,Eichhorn:2016esv}.}

Despite of their common root given by the partition sum \eqref{partfct}, a systematic link between results obtained within the CDT program and the FRGE approach is still missing. While the spectral dimension of the resulting quantum spacetimes have been compared in \cite{Reuter:2011ah}, little is known about the relation of the two formulations. On this basis, the present work devises an FRGE study which incorporates all the essential features underlying the Monte Carlo simulations carried out within CDT. The natural continuum analogue of the foliation structure imposed on the microscopic spacetimes studied within CDT is the Arnowitt-Deser-Misner (ADM) formulation reviewed, e.g., in \cite{PoissonBook}. In this formalism spacetime is built up from a stack of spatial hypersurfaces $\Sigma_\tau$ on which the time-variable $\tau$ is constant. These hypersurfaces are welded together such that they fill the entire spacetime. The resulting preferred ``time'' direction obtained in this way plays a similar role as the causal structure implemented in CDT. 

An FRGE tailored to the ADM-formalism has been constructed in \cite{Manrique:2011jc,Rechenberger:2012dt} and we will use this framework in the sequel.\footnote{For applications of this formalism to Ho\v{r}ava-Lifshitz gravity \cite{Horava:2009uw} see \cite{Contillo:2013fua,D'Odorico:2014iha}.}
This construction makes manifest use of the background field formalism. Since most CDT simulations restrict the geometries contributing to \eqref{partfct} to be of topology $S^1 \times S^2$ or $S^1 \times S^3$,
we evaluate the flow equation on a background geometry given by $S^1 \times S^d$ where the (intrinsic) curvature of $S^d$ is a free parameter. Moreover, the flow is projected onto Einstein-Hilbert action which provides the weight of the partition sum \eqref{partfct} in the CDT framework.

Our work is complementary to the recent investigation \cite{Biemans:2016rvp,Biemans:2017zca} in the sense that it uses a different background topology. It also provides a detailed analysis on how the flow is influenced by integrating over different classes of spatial fluctuations and under a change of the regulator scheme. As a main result, we find that all cases studied in this paper admit a NGFP suitable for Asymptotic Safety. The phase diagrams obtained from integrating the flow equations are  strikingly similar to the ones found for background topology $S^1 \times T^d$ \cite{Biemans:2016rvp,Biemans:2017zca}. In particular, we show that there are specific combinations of parametrizing the metric fluctuations and regulating the flow equation which realizes the double-fixed point scenario found in \cite{Biemans:2016rvp} \emph{in four spacetime dimensions}. In this case the RG trajectory realized by Nature, as described in \cite{Reuter:2004nx}, is well-defined on all length scales. The mechanism underlying the completion of the RG trajectories in the deep infrared is closely related to the proposal of ``erasing the cosmological constant through a gravitational instability'', recently made in \cite{Wetterich:2017ixo}.

The rest of the work is organized as follows. Sect.\ \ref{sect.2} introduces the essential elements of the ADM-formalism together with the corresponding FRGE. The beta functions governing the flow of Newton's coupling and the cosmological constant are constructed in Sect.\ \ref{sect.3} and their properties are analyzed in Sect.\ \ref{sect.4}. We close with a brief discussion of our findings in Sect.\ \ref{sect.5}. Technical details about the background geometry, the structure of the flow equation, and the evaluation of the operator traces are provided in App.\ \ref{App.A}, App.\ \ref{App.B}, and App.\ \ref{App.C}, respectively.

\section{Renormalization group flows in the ADM-formalism}
\setcounter{equation}{0}
\label{sect.2}
Our construction of the gravitational renormalization group (RG) flow is based on the FRGE for the effective average action \cite{Wetterich:1992yh,Morris:1993qb,Reuter:1993kw} tailored to the Arnowitt-Deser-Misner (ADM) formulation  \cite{Manrique:2011jc,Rechenberger:2012dt}. This section summarizes the central points of the construction.

\subsection{Parametrization of the fluctuation fields}
The ADM formalism decomposes the spacetime metric $g_{\mu\nu}$ into a lapse function $N(\tau,y)$, a shift vector $N_i(\tau,y)$ and a metric $\sigma_{ij}(\tau,y)$. The later measures distances on the spatial slices $\Sigma_\tau$ defined by $\tau = const$. For Euclidean signature this decomposition is given by
\be\label{fol1}
ds^2 = g_{\mu\nu} \, dx^\mu dx^\nu
=  N^2 d\tau^2 +  \sigma_{ij} \, (dy^i + N^i d \tau)
(dy^j +  N^j d \tau) \, .
\ee
At the level of the metric tensor, this entails
\be\label{metcomp}
g_{\mu\nu} = \left(
\begin{array}{cc}
	N^2 + N_i N^i \; \;  & \;  \; N_j \\
	N_i &  \sigma_{ij}
\end{array}
\right) \, , \quad
g^{\mu\nu} = \left(
\begin{array}{cc}
	\frac{1}{N^{2}} \; \;  & \;  \; -  \frac{N^j}{ N^{2}}  \\
	-  \frac{N^i}{ N^{2}}  	 \; \;  & \;  \;  \sigma^{ij} +  \, \frac{ N^i \,  N^j}{ N^{2}}
\end{array}
\right) \, .
\ee
An infinitesimal coordinate transformation acts on the spacetime metric via $\delta g_{\alpha\beta} = \cL_v \, g_{\alpha\beta}$ where $\cL_v$ is the Lie derivative. This transformation induces the transformation law for the component fields
\be\label{eq:gaugeVariations}
\begin{split}
	\delta N &= \p_\tau (f N ) + \zeta^k \p_k N  - N N^i\p_i f \, , \\
	\delta  N_i &= \partial_\tau( N_i f) + \zeta^k\partial_k  N_i +  N_k\partial_i\zeta^k
	+ \sigma_{ki}\partial_\tau \zeta^k
	+  N_k  N^k\partial_i f  +  N^2\partial_i f \, , \\
	\delta\sigma_{ij} &= f \, \p_\tau \sigma_{ij} + \zeta^k \, \p_k \sigma_{ij} + \sigma_{jk} \, \p_i\zeta^k + \sigma_{ik} \, \p_j\zeta^k + N_j\, \p_i f + N_i\p_j f  \, ,
\end{split}
\ee
where the vector $v^\alpha$ has been decomposed into a time-component $f$ and a vector tangent to the spatial slice $\zeta^i$.

The construction of the flow equation for the ADM formalism uses the background field method. The quantum fields $N, N_i, \sigma_{ij}$ are decomposed into a fixed (but arbitrary) background $\Nb, \Nb_i, \sib_{ij}$ and fluctuations around this background $\Nh, \Nh_i, \sih_{ij}$. For the lapse function and the shift vector, we resort to a linear split
\be\label{lapseshiftdec}
N = \Nb + \Nh \, , \qquad N_i = \Nb_i + \Nh_i \, .
\ee
The fluctuations of $\sigma_{ij}$ may be parametrized either through a linear or an exponential split
\be\label{flucpara}
\begin{split}
\mbox{linear:} \qquad & \sigma_{ij} = \sib_{ij} + \sih_{ij} \, , \\
\mbox{exponential:} \qquad & \sigma_{ij} = \sib_{il} \left[ \, e^{\sih} \, \right]^l{}_j\, .
\end{split}
\ee
Here indices are raised and lowered with the background metric. Essentially, the choice of split \eqref{flucpara} determines the type of fluctuations admissible in the construction: the exponential split guarantees that $\sigma_{ij}$ and $\sib_{ij}$ have the same signature while in the linear split the fluctuations may change the signature of $\sigma_{ij}$. The exponential split of the spatial metric in the ADM decomposition then has the same effect as in the covariant construction
\cite{Gies:2015tca,Ohta:2016npm,Ohta:2016jvw,Falls:2017cze}: in both cases fluctuations can not change the signature of the metric.

At the level of the spacetime metric, the exponential split \eqref{flucpara} ensures that the signature of the time direction and spatial metric remains unchanged, independently of the value of the fluctuation fields. This can be seen from computing the determinant of $g_{\mu\nu}$. Applying the method of Schur complements yields
\be
\det\left( g_{\mu\nu}\right) = N^2 \, \det \left( \sigma_{ij} \right) \, .
\ee
Here $N^2$ is positive by construction and the exponential split ensures that  $\sigma_{ij}$ has the same signature as $\sib_{ij}$. Thus the ADM formalism in the exponential parametrization constitutes a refinement of the standard exponential parametrization by restricting the quantum fluctuations to the set which conserves the signature of the spatial and time part of the spacetime metric independently.

In terms of practical computations, it is convenient to combine the linear and exponential splits according to
\be\label{combsplit}
\sigma_{ij} \simeq \sib_{ij} + \sih_{ij} + \tfrac{\alpha}{2} \, \sih_{ik} \, \sib^{kl} \, \sih_{lj} + \ldots \, .
\ee
The parameter $\alpha$ takes the value $\alpha = 0$ for the linear and $\alpha = 1$ for the exponential split. The dots represent terms containing cubic and higher powers of the fluctuation fields. Since these terms will not contribute to the present computation we refrain from giving their explicit structure at this stage.

\subsection{The functional renormalization group equation}
The scale-dependence of coupling constants can conveniently be obtained from the FRGE for the effective average action $\Gamma_k$ \cite{Wetterich:1992yh,Morris:1993qb,Reuter:1993kw,Reuter:1996cp}. Besides the gravitational action, $\Gamma_k$ also contains suitable gauge-fixing and ghost terms
\be\label{eaa}
\Gamma_k[\hat{\chi}; \bar{\chi} ] = \Gamma_k^{\rm grav}[\hat{\chi}; \bar{\chi} ] + \Gamma_k^{\rm gauge-fixing}[\hat{\chi}; \bar{\chi} ] + \Gamma_k^{\rm ghost}[\hat{\chi}; \bar{\chi} ] \, .
\ee
Here $\hat{\chi}$ and $\bar{\chi}$ denote the collection of fluctuation fields and background fields, respectively. The central property of $\Gamma_k$ is that its dependence on the RG scale $k$ is governed by the formally exact RG equation
\be\label{FRGE}
k \p_k \Gamma_k = \frac{1}{2} \, {\rm Tr} \left[ \left( \Gamma_k^{(2)} + \cR_k \right)^{-1} \, k \p_k \cR_k \right] \, .
\ee
Here $\Gamma_k^{(2)}$ is the second variation of $\Gamma_k$ with respect to the fluctuation fields and the trace indicates an integration over loop momenta. 
The regulator $\cR_k$ provides a $k$-dependent mass term for the fluctuation modes with momenta $p^2 \ll k^2$ and vanishes for $p^2 \gg k^2$. 
 In the propagator $\left( \Gamma_k^{(2)} + \cR_k \right)^{-1}$, the regulator suppresses the contribution of fluctuations with momenta $p^2 \ll k^2$ to the trace. The term $k \p_k \cR_k$ in the numerator ensures that  fluctuations with $p^2 \gg k^2$ do not contribute to the trace. As a consequence the right-hand-side of eq.\ \eqref{FRGE} is finite. Moreover, the flow of $\Gamma_k$ is driven by fluctuations whose momenta are comparable to the RG scale $k$.

The FRGE realizes several welcome features. Firstly, vertices extracted from $\Gamma_k$ include quantum corrections resulting from integrating out fluctuations with momenta $p^2 \gtrsim k^2$. Thus $\Gamma_k$ provides a one-parameter family of effective descriptions of physics at the scale $k$. This realizes Wilson's idea of renormalization. Secondly, the FRGE may be used to study the RG flow and phase diagram of a theory without specifying an initial or fundamental action. This feature is particularly relevant in the context of Asymptotic Safety where the fundamental action is unknown a priori and arises as a fixed point of the flow.
Such fixed points may be visible already in relatively simple projections of the FRGE. In the next section, we will utilize this feature and make a specific ansatz for the effective average action \eqref{eaa} in order to study RG flow of Newton's coupling and the cosmological constant in a setting tailored to CDT.

\section{Einstein-Hilbert truncation on $S^1 \times S^d$}
\setcounter{equation}{0}
\label{sect.3}
We now use the FRGE \eqref{FRGE} to construct the beta functions governing the flow of Newton's coupling and the cosmological constant on a background topology $S^1 \times S^d$.
\subsection{Ansatz for the effective average action}
We approximate the gravitational part of $\Gamma_k$ by the Euclidean Einstein-Hilbert action. In terms of the ADM fields the resulting action is given by
\be\label{EHaction}
\Gamma_k^{\rm grav} = \frac{1}{16 \pi G_k} \int d\tau d^dy \, N \sqrt{\sigma} \left[ K_{ij}  K^{ij} - K^2 - R + 2 \Lambda_k \right] \, .
\ee
Here
\be\label{Kext}
K_{ij} \equiv \frac{1}{2 N} \left( \p_\tau \sigma_{ij} - D_i N_j - D_j N_i \right) \, , \qquad K \equiv \sigma^{ij} K_{ij} \, ,
\ee
denotes the extrinsic curvature and $R$ is the intrinsic curvature constructed from $\sigma_{ij}$. The ansatz comprises two scale-dependent couplings, Newton's coupling $G_k$ and the cosmological constant $\Lambda_k$.

At this stage, it is convenient to make an explicit choice for the background fields. In the present work, we will choose a class of backgrounds with topology $S^1 \times S^d$. Explicitly, we take
\be\label{background}
\Nb = 1 \, , \qquad \Nb_i = 0 \, , \qquad \sib_{ij}(\tau,y) = \sib_{ij}^{S^d}(y)
\ee
where $\sib_{ij}^{S^d}$ is a one-parameter family of metrics on the sphere $S^d$ with radius $r$ and independent of $\tau$. The curvature tensors for this background geometry then satisfy
\be
\Kb_{ij} = 0 \, , \quad \Rb_{ijkl} = \tfrac{\Rb}{d(d-1)} \left( \sib_{ik} \sib_{jl} - \sib_{il} \sib_{jk} \right) \, , \quad \Rb_{ij} = \tfrac{1}{d} \sib_{ij} \Rb \, ,
\ee
with $\Rb$ being constant. A key property of this background is the existence of a global Killing vector field $\p_\tau$. This feature allows to perform a Wick rotation without generating a complex background geometry. Denoting expressions in Euclidean and Lorentzian signature by subscripts $E$ and $L$, the corresponding analytic continuation is given by
\be\label{Wickrot}
\tau_E \rightarrow - i \tau_L \, , \qquad N^i_E \rightarrow i N^i_L \, .
\ee

At this stage it is instructive to verify that this background is sufficient to disentangle the flow of $G_k$ and $\Lambda_k$. For this purpose, we take the $k$-derivative of the ansatz \eqref{EHaction} and subsequently set the fluctuation fields to zero
\be
\left. k \, \p_k \Gamma_k^{\rm grav} \right|_{\hat{\chi} = 0} = k \p_k \left(
\frac{1}{16 \pi G_k} \int d\tau d^dy \sqrt{\sib} \left[ - \Rb + 2 \Lambda_k \right]
 \right) \, .
\ee
This indicates that on the background \eqref{background} the flow of Newton's coupling can be constructed from the coefficients multiplying the intrinsic background curvature while the beta function for the cosmological constant is encoded in the volume terms appearing on the left- and the right-hand-side of the FRGE. Thus it suffices to keep track of these two terms in the following.

The gravitational part of the effective average action has to be complemented by a suitable gauge fixing and ghost action. Following the strategy \cite{Biemans:2016rvp}, we use the gauge freedom in such a way that all fluctuation fields including the lapse function and the shift vector acquire a relativistic dispersion relation. Moreover, terms which are of the form $E \sqrt{\vec p^{\, \, 2}}$ are consistently eliminated from the Hessian $\Gamma_k^{(2)}$. These two conditions actually fix the choice of gauge uniquely
\be\label{Sgf}
\Gamma_k^{\rm gf} = \frac{1}{32 \pi G_k} \int d\tau d^dy \sqrt{\sib} \, \left[
F^2 + F_i  \, \sib^{ij} \, F_j \right] \, .
\ee
The $F$ and $F_i$ are linear in the fluctuation fields and read
\be\label{gaugefixing}
\begin{split}
F = & \, \left[ \partial_\tau\hat{N}+ \bar{D^i}\hat{N_i} - \tfrac{1}{2} \partial_\tau\hat{\sigma} \right] \\
F_i = & \, \left[\partial_\tau\hat{N_i}-\bar{D}_i\hat{N} - \tfrac{1}{2} \bar{D}_i\hat{\sigma} + \bar{D^j}\hat{\sigma}_{ij} \right] \, ,
\end{split}
\ee
where $\sih \equiv \sib^{ij} \sih_{ij}$ and $\Db_i$ denotes the covariant derivative constructed from $\sib_{ij}$. The gauge fixing \eqref{gaugefixing} can be derived by adapting the harmonic gauge condition to the specific class of backgrounds and specifying the parametrization of the fluctuation fields to the one generated by the ADM decomposition. The action of the Faddeev-Popov ghosts is then constructed in the standard way. It comprises one pair of scalar ghosts $\bar{c},c$ and one pair of spatial vector ghosts $\bar{b}^i, b_i$. Restricting to terms quadratic in the fluctuation fields, the action reads
\be\label{ghosts}
\Gamma_k^\mathrm{gh}= \int d\tau d^dy \sqrt{\sib} \, \big[ \, \bar{c} \, \Delta \, c + \bar{b}^i \, \left(\Delta - \tfrac{\Rb}{d} \right) \, b_i  \big] \, ,
\ee
where $\Delta \equiv - \p_\tau^2 - \sib^{ij} \Db_i \Db_j$ is the $D$-dimensional Laplace operator constructed from the background spacetime.

In order to compute the propagator $(\Gamma_k + \cR_k)^{-1}$, it is useful to perform a transverse traceless decomposition of the fluctuation fields which is adapted to the background. The shift vector is decomposed into a transverse vector $u_i$ and a scalar $B$
\be\label{dec1}
\Nh_i = u_i + \Db_i B \, , \qquad \Db^i u_i = 0 \, . 
\ee 
For the fluctuations of the spatial metric, we resort to the standard transverse-traceless decomposition of a symmetric tensor,
\be\label{dec2}
\hat{\sigma}_{ij} = h_{ij} + \Db_i v_j + \Db_j v_i + \left(\Db_i \Db_j-\tfrac{1}{d}\, \bar{\sigma}_{ij} \, \Db^2\right)\psi + \tfrac{1}{d} \, \bar{\sigma}_{ij}h
\ee
where the component fields are subject to the constraints 
\be
\Db^i h_{ij} = 0, \qquad \bar{\sigma}^{ij} h_{ij} = 0, \qquad \Db^i v_i = 0, \qquad \sib^{ij} \hat{\sigma}_{ij} = \sih = h \, .
\ee
The Jacobians coming from these decompositions are absorbed into the momentum-dependent field redefinition \eqref{redef}. The matrix elements of $\Gamma_k^{(2)}$ with respect to these component fields are computed in App.\ \ref{App.B} and summarized in Tab.\ \ref{Tab.3}. The result shows that the field decomposition diagonalizes $\Gamma_k^{(2)}$ in field space, apart from the scalar sector containing the two fields $\Nh$ and $h$.  

The final ingredient required in the evaluation of the FRGE is the regulator $\cR_k$. On the background \eqref{background} the Hessian $\Gamma_k^{(2)}$ is matrix valued in field space. From Tab.\ \ref{Tab.3} one finds that the typical matrix element has the structure
\be\label{structure}
\left[ \Gamma_k^{(2)} \right]_{ab} = \left(32 \pi G_k\right)^{-s} \, \left[ \Delta + q \, \Rb + \ldots \right]_{ab} \, ,
\ee
where $s$ takes the values $0,1$ and the dots represent a possible contribution from the cosmological constant. Moreover, $q$ is a fixed, $d$-dependent numerical coefficient which depends on the field indices $a,b$. For example, the ghost action \eqref{ghosts} leads to $s=0$ and  $q_{\bar{c}c} = 0$ and $q_{\bar{b}b} = -1/d$. Based on the structure \eqref{structure} there are two natural choices for a coarse graining operator. Following the nomenclature introduced in \cite{Codello:2008vh}, we define
\be\label{regscheme}
\begin{array}{ll}
{\rm Type \; I}: \qquad & \Box \equiv \Delta \, , \\[1.0ex]
{\rm Type \; II}: \qquad & \Box \equiv \Delta + q \, \Rb \, . 	
\end{array}
\ee
For notational convenience, the two regularization schemes \eqref{regscheme} are combined by introducing a  parameter $r$ via
\be
\Box \equiv \Delta + r \, q \, \Rb \, .
\ee
Setting $r=0$ or $r=1$ then corresponds to a regulator scheme of Type I and Type II, respectively. The matrix-elements of $\cR_k$ are then taken as operator-valued functions depending on the coarse-graining operator. Their explicit form can be obtained by the replacement rule
\be\label{cutoff}
\Box \mapsto P_k \equiv \Box + R_k(\Box) \,
\ee
where $R_k(z)$ is a scalar profile function. The parametrization of the fluctuation fields combined with the specific choices for the coarse-graining procedure, which are considered in the following, are summarized in Tab.\ \ref{Tab.schemes}. At this stage, we have all the ingredients for evaluating the flow equation for the ansatz \eqref{EHaction}.
\begin{table}[t!]
	\renewcommand{\arraystretch}{1.3}
	\begin{center}
		\begin{tabular}{|c|c||c|c|c|} \hline
	\hspace*{3mm}	metric fluctuations \hspace*{3mm} & \hspace*{3mm} regulator \hspace*{3mm} & \hspace*{5mm} $\alpha$ \hspace*{5mm} & \hspace*{5mm} $r$ \hspace*{5mm} & \hspace*{5mm} $q_{\rm off-diag}$ \hspace*{5mm} \\ \hline \hline
		\multirow{4}{*}{linear} & I & $0$ & $0$ & $0$ \\ \cline{2-5}
		& II$_0$ & $0$ & $1$ & $0$ \\ \cline{2-5} 
		& II$_1$ & $0$ & $1$ & $\frac{d-2}{d}$ \\ \cline{2-5} 
		& II$_2$ & $0$ & $1$ & $\frac{d-4+\alpha}{d}$ \\ \hline
		\multirow{4}{*}{exponential} & I & $1$ & $0$ & $0$ \\ \cline{2-5}
		& II$_0$ & $1$ & $1$ & $0$ \\ \cline{2-5} 
		& II$_1$ & $1$ & $1$ & $\frac{d-2}{d}$ \\ \cline{2-5} 
		& II$_2$ & $1$ & $1$ & $\frac{d-4+\alpha}{d}$ \\ \hline
		\end{tabular}
		\caption{\label{Tab.schemes} Parameter sets used in analyzing the dynamics encoded in the beta functions \eqref{betafcts}. The value $q_{\rm off-diag}$ specifies the endomorphism in the $\Nh$-$h$--sector, where the coarse-graining operator is then given by $\Box = \Delta + q_{\rm off-diag} \Rb$.}
	\end{center}
\end{table}

We remark that for a Type II regulator scheme the coarse graining operators $\Box$ are not necessarily positive semi-definite. Their explicit spectrum can be constructed from the eigenvalues of the Laplacian on the $d$-sphere listed, e.g., in \cite{Lauscher:2001ya}. In this way one finds that setting $r=1$ implies that $\Box$ has negative eigenvalues when acting on the constant $\psi$-mode and the two lowest eigenmodes in the $B$-sector if $d=3$. In addition the linear split combined with the II$_2$ regularization scheme leads to negative eigenvalues in the $\Nh$-$h$-sector. The mode count for the vector ghosts is identical to the $B$-$u_i$ sector. The negative eigenvalues of $\Box$ occurring in the Type II case then suggest that the Type I regularization procedure may be preferred. The possibility for adjusting the spectrum of the coarse-graining operator by including a suitable endomorphism component may be used to implement conditions similar to the ``equal lowest eigenvalue scheme'' advocated in \cite{Demmel:2014hla}. While it would be desirable to have a more complete understanding of the regulator dependence in the present case, we limit ourselves to the analysis of the cases introduced in Tab.\ \ref{Tab.schemes}.

We close this subsection with the following remark. In the companion paper \cite{Biemans:2016rvp} the setup \eqref{EHaction} was used to construct the beta functions of $G_k$ and $\Lambda_k$ on a Euclidean Friedmann-Robertson-Walker background. In this case the background geometry is characterized by $\Rb = 0$ while one has a non-vanishing extrinsic curvature $\Kb_{ij}$. The beta function for the Newton's coupling is then read off from the extrinsic curvature terms. At the level of classical general relativity
the two incarnations of $G_k$ related to the extrinsic and extrinsic curvature terms coincide due to diffeomorphism invariance of the Einstein-Hilbert action. At the level of the FRGE it is expected that the two projection schemes may lead to (slightly) different results. Firstly, introduction of the regulator $\cR_k$ may break the full diffeomorphism group to a subgroup so that the two projections actually construct the flow of two different coupling constants. Moreover, working on different backgrounds may result in different organization schemes for the fluctuation modes, indicating that modes integrated out at a certain scale $k$ could be different in the two settings. The setting of this paper provides an ideal testing ground for obtaining a quantitative understanding of these effects.

\subsection{Beta functions}
The beta functions governing the scale dependence of $G_k$ and $\Lambda_k$ are constructed in App.\ \ref{App.C}. For conciseness, we limit ourselves to the expression obtained from setting the endomorphism piece in the scalar sector spanned by $\Nh$ and $h$ to zero. Moreover, all threshold functions are evaluated with a Litim-type regulator \eqref{thresholdfcts}. The result is conveniently expressed in terms of the dimensionless quantities
\be\label{dimless}
\eta \equiv (G_k)^{-1} \p_t \, G_k \, , \qquad
\lambda_k \equiv \Lambda_k \, k^{-2} \, , \qquad
g_k \equiv G_k \, k^{d-1} \, ,
\ee
where $\eta$ is the anomalous dimension of Newton's coupling.
The scale dependence of the dimensionful couplings is then governed by
the beta functions
\be\label{betafcts}
k \p_k g_k = \beta_g(g_k, \lambda_k) \, , \qquad k \p_k \lambda_k = \beta_\lambda(g_k, \lambda_k) \, .
\ee
Defining
\be\label{Bdet}
B_{\rm det} \equiv d-1 - (3d-2+ \alpha) \lambda + 2 d \lambda^2
\ee
one has
\be\label{betaexp}
\begin{split}
\beta_g = & \, (d-1+\eta) \, g \, , \\
\beta_\lambda = & \, (\eta - 2) \lambda 
+ \tfrac{g}{(4\pi)^{(d-1)/2}} \Big[  -\tfrac{4 (d+1)}{\Gamma((d+3)/2)} 
\\ & 
\quad 
+ \left( d + \tfrac{d^2+d-2}{2(1-(2-\alpha)\lambda)} + \tfrac{2(d-1) - (3d - 2 + \alpha ) \lambda}{B_{\rm det}} \right) \left( \tfrac{2}{\Gamma((d+3)/2)}
 - \tfrac{\eta}{\Gamma((d+5)/2)} \right) 
 \Big] \, . 
 \end{split}
\ee
The anomalous dimension of Newton's coupling takes the form
\be\label{eta}
\eta = \frac{16 \pi g \, B_1(\lambda)}{(4\pi)^{(d+1)/2} + 16 \pi g \, B_2(\lambda)} \, . 
\ee
The functions $B_1$ and $B_2$ depend on $\lambda$ as well as the parameters $d,r,\alpha$. The terms appearing in these expressions are conveniently organized in terms of the contributions found in App.\ \ref{App.C2}
\be
B_{a} = B_{a,1} + B_{a,2} + B_{a,3} + B_{a,4} + B_{a,5} \, , \qquad a =1,2 \, . 
\ee 
The explicit expressions for the building blocks are
\be
\begin{split}
B_{1,1}	= & \, - \tfrac{d^3+3d^2+20d-6 + 6 r (d-1)^2}{12 \, d \, \Gamma((d+3)/2)} + \tfrac{\delta_{2,d}}{d \, \Gamma((d+1)/2)}\, , \\[1.2ex]
B_{1,2}	= & \, \tfrac{d^4 - 15 d^2 - 22d + 12
	  - 3 r \left( d^2 ( d-3) (1 + d) (2-\alpha) + 4 d \, (3 - 2 \alpha) - 4 \, (3 - \alpha )  \right) + 3\left( d^3+3d^2+6d-4 \right) \delta_{2,d}}{12 \,d\, (d-1) \, (1 - (2-\alpha) \lambda) \, \Gamma((d+1)/2)} \, , \\[1.2ex]
B_{1,3}	= & \, (r-1) \, \tfrac{ d^2 (d^2-2d-3)(2-\alpha) + 4 d (3 - 2 \alpha)  - 4 (3 - \alpha)
		}{4 d (d-1) (1-(2-\alpha)\lambda)^2  \, \Gamma((d+3)/2)} \, , \\[1.2ex]
B_{1,4}	= & \, \tfrac{2(d-1) - (3d - 2 + \alpha ) \lambda}{6 \, B_{\rm det} \, \Gamma((d+1)/2)}  \, , \\[1.2ex]
B_{1,5}	= & \, - \tfrac{(d-2)}{2d} \, \tfrac{ (d-1) ( 3 d -4 + \alpha ) -8d(d-1) \lambda + 2 d (3 d + \alpha) \lambda^2 }{ B_{\rm det}^2 \, \Gamma((d+3)/2)}  \, ,
\end{split}
\ee
and
\be
B_{2,1}	= \tfrac{d^3 + 3 d^2 + 6 d -6  + 6 r (d+1)^2}{24d \,  \Gamma((d+5)/2)} + \tfrac{\delta_{2,d}}{2 d \, \Gamma((d+3)/2)} \, . 
\ee
The remaining coefficients in $B_2$ are proportional to their $B_1$ counterparts
\be
B_{1,2} = (d+1) B_{2,2} \, , \; \;
B_{1,3} = (d+1) B_{2,3} \, , \; \;
B_{1,4} = (d+3) B_{2,4} \, , \; \;
B_{1,5} = (d+3) B_{2,5} \, . 
\ee
This result completes the derivation of the beta functions for $g_k$ and $\lambda_k$ on the background topology $S^1 \times S^d$. Notably $\beta_\lambda$ is independent of the endomorphism parameter $r$ but retains information on the parametrization of the metric fluctuations. For $\alpha = 0$, it agrees with the flow of the cosmological constant obtained on a $S^1 \times T^d$-background \cite{Biemans:2017zca}. The beta functions \eqref{betaexp} are the main result obtained in this section.

\section{Properties of the renormalization group flow}
\setcounter{equation}{0}
\label{sect.4}
The beta functions derived in the previous section explicitly retain information on the parametrization of the fluctuation fields, encoded in the parameter $\alpha$, and the choice of regularization scheme, parametrized by $r$. Typically, these parameters have distinguished values. In the following subsection, we investigate how these choices affect the flow of $g_k$ and $\lambda_k$. Throughout the discussion we will limit ourselves mostly to the case $d=3$, corresponding to a four-dimensional spacetime.
\subsection{Fixed point structure and phase diagrams}
\label{sect.41}
The beta functions \eqref{betafcts} constitute a system of autonomous coupled non-linear differential equations. In order to understand the dynamics of the system, it is useful to first determine its fixed points and singularity structure.

Singularities in the beta functions \eqref{betaexp} can be traced back to two sources. First, there are loci in the $g$-$\lambda$-plane where the threshold functions diverge. The location of these lines depends on $\alpha$ and is independent of $r$. Evaluating the roots of eq.\ \eqref{Bdet}, one finds
\be\label{fixedsing}
\begin{tabular}{lll}
$\alpha = 0$: \qquad \quad & $\lambda_1^{\rm sing} = \tfrac{1}{2}$\, ,  \qquad & $\lambda_2^{\rm sing} = \tfrac{d-1}{d}$ \, , \\[1.2ex]
$\alpha = 1$: \qquad \quad & $\lambda_1^{\rm sing} = \tfrac{d-1}{2d}$\, ,  \qquad & $\lambda_2^{\rm sing} = 1$ \, . 
\end{tabular}
\ee
All singular lines are independent of $g$ and located at $\lambda > 0$. Notably, the exponential parametrization moves the singular line $\lambda_1^{\rm sing}$ closer to the origin. The singularity $\lambda_1^{\rm sing}$ is the counterpart of the gravitational instability discussed in Ref.\ \cite{Wetterich:2017ixo}. Besides these fixed singularities, there are also lines in the $g$-$\lambda$-plane where the anomalous dimension of Newton's coupling diverges. In this case the denominator in eq.\ \eqref{eta} vanishes. Exploiting the $B_2(\lambda)$ is independent of $g$, this line is conveniently described by a parametrized curve obtained by equating the denominator of $\eta$ to zero and solving for $g^{\rm sing}$:
\be\label{etasing}
\eta^{\rm sing}: \qquad \quad g^{\rm sing} = \frac{2^{(d-3)} \,  \pi^{(d-1)/2}}{B_2(\lambda)} \, . 
\ee
The position of this singular line depends on both parameters $r$ and $\alpha$. 

\begin{figure}[t!]
	\begin{center}
		\includegraphics[width=0.48\textwidth]{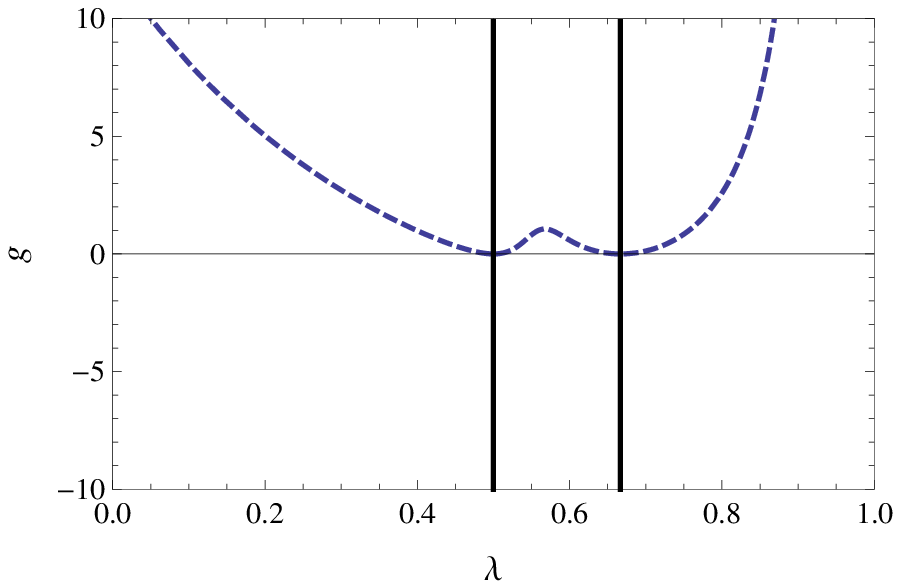} \;
		\includegraphics[width=0.48\textwidth]{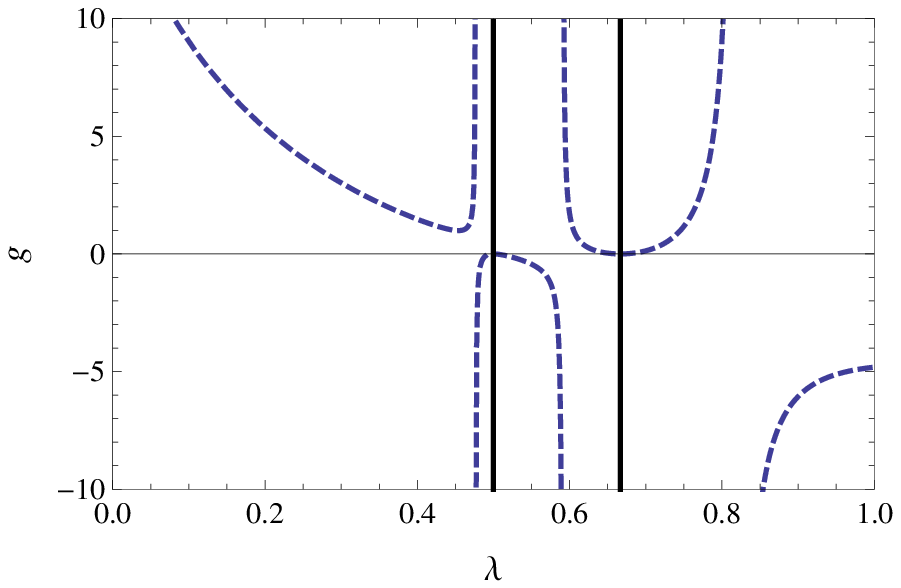} \\
		\caption{\label{singloci} Singular loci of the beta functions \eqref{betaexp} for the Type I, linear (left) and Type II$_0$, linear (right) setup in $d=3$. The black, solid lines show the fixed singularities \eqref{fixedsing} while the divergence of $\eta$ is given by the dashed blue line. The two diagrams illustrate the two prototypical cases where the fixed singularity is screened (left) or unscreened (right) by $\eta^{\rm sing}$.}
	\end{center}	
\end{figure}

The singular lines \eqref{fixedsing} and \eqref{etasing} are shown in Fig.\ \ref{singloci}. At this point it is useful to distinguish between the two qualitatively different scenarios. 
Focusing on the region where $g>0$ one inevitably encounters a singular line when moving to positive values of $\lambda$. In the first setting, this singular locus is associated with a divergence of the anomalous dimension $\eta$ given by \eqref{etasing}. The prototypical singularity structure for this case is shown in the left diagram of Fig.\ \ref{singloci}. It is realized for the exponential parametrization ($\alpha=1$) and the linear parametrization ($\alpha = 0$) combined with a Type I and Type II$_2$ regularization scheme. In the second setting the locus $\eta^{\rm sing}$ has a pole located before the first fixed singularity. As a consequence the line $\lambda_1^{\rm sing}$ is not entirely shielded by the divergence of $\eta$. This scenario is realized for the linear parametrization $\alpha = 0$ with the Type II$_0$ and Type II$_1$ regulator. It will be shown below that the different singularity structures lead to qualitatively different low-energy behaviors of the RG flow in the region $\lambda > 0$. 

Subsequently, we analyze the fixed point structure of the beta functions.
 At a fixed point $(g_*, \lambda_*)$ the beta functions vanish by definition
\be\label{deffp}
\beta_g(g_*,\lambda_*) = 0 \, , \quad \beta_\lambda(g_*,\lambda_*) = 0 \, . 
\ee
In the vicinity of a fixed point the properties of the RG flow can be studied by linearizing the beta functions. The dynamics of the linearized system is encoded in the stability matrix ${\bf B}_{ij} = \p_{g_j} \beta_{g_i}|_{g = g_*}$. The stability coefficients $\theta$ are defined as minus the eigenvalues of ${\bf B}_{ij}$. For eigendirections with $\theta > 0$ the solutions are dragged into the fixed point for $k \rightarrow \infty$ while eigendirections with $\theta < 0$ repel the flow in this limit.
\begin{table}[t!]
	\renewcommand{\arraystretch}{1.3}
	\begin{center}
		\begin{tabular}{|c|c||c|c|c||c|c|}
			\hline
			\hspace*{2mm} fluctuations  \hspace*{2mm} & \hspace*{2mm} regulator \hspace*{2mm} & \hspace*{4mm} $g_*$ \hspace*{4mm} & \hspace*{4mm} $\lambda_*$ \hspace*{4mm} & \hspace*{4mm} $g_* \lambda_*$ \hspace*{4mm} &   \multicolumn{2}{c|}{\hspace*{10mm} $\theta_{1,2}$ \hspace*{10mm}}  \\ \hline \hline
			\multirow{10}{*}{linear} & \multirow{2}{*}{Type I}  & $0.901$ & $0.222$ & $0.200$ & \multicolumn{2}{c|}{$1.432 \pm 2.586 i $} \\
			& & $-$ & $-$ & $-$ & $-$ & $-$ 
			\\ \cline{2-7}  
			& \multirow{2}{*}{Type $\mathrm{II}_0$} &  $0.896$ & $0.203$ & $0.182$ & \multicolumn{2}{c|}{$1.545 \pm 2.032 i $} \\ 
			& & $0.342$ & $	0.438$ & $0.150$ & $2.774$ & $-23.89$ \\ \cline{2-7} 
			& \multirow{2}{*}{Type $\mathrm{II}_1$} &  $0.879$ & $0.182$ & $0.160$ & \multicolumn{2}{c|}{$1.765 \pm 1.787 i $} \\ 
			& & $0.510$ & $	0.400$ & $0.204$ & $3.016$ & $-13.28$ \\ \cline{2-7} 
			& \multirow{2}{*}{Type $\mathrm{II}_2$} &  $0.901$ & $0.222$ & $0.200$ & \multicolumn{2}{c|}{$1.329 \pm 2.332 i $} \\ 
			& & $-$ & $	-$ & $-$ & $-$ & $-$ \\ \cline{2-7}
			& \multirow{2}{*}{optimized} & $0.900$ & $0.230$ & $0.207$ & \multicolumn{2}{c|}{$1.414 \pm 2.941 i $} \\ 
			& & $-$ & $	-$ & $-$ & $-$ & $-$
			\\ \hline \hline
			\multirow{3}{*}{exponential} &	Type I & $1.049$ & $0.249$ & $0.262$ & \multicolumn{2}{c|}{$0.444 \pm 4.041 i $} \\ \cline{2-7}
			& Type $\mathrm{II}_0$ & $1.050$ & $0.249$ & $0.261$ & \multicolumn{2}{c|}{$0.342 \pm 3.855 i $} \\ \cline{2-7}
			& Type $\mathrm{II}_2$ & $1.050$ & $0.249$ & $0.261$ & \multicolumn{2}{c|}{$0.342 \pm 3.855 i $} \\ \hline \hline
		\end{tabular}
		\caption{\label{tab.1} NGFPs of the beta functions \eqref{betaexp} evaluated for the linear split ($\alpha = 0$) and the exponential split ($\alpha = 1$) and regulators of Type I and Type II, respectively. The NGFP obtained from the optimization procedure shown in Fig.\ \ref{Fig.opt} is listed with the label ``optimized''.}
	\end{center}
\end{table}
All implementations of the beta functions posses a Gaussian fixed point (GFP) located in the origin. This fixed point corresponds to the free theory and its critical exponents are given by the mass dimension of the dimensionful couplings. Besides the GFP the beta functions also possess non-Gaussian fixed points (NGFPs). Limiting to the physically interesting region with $g >0$ located to the left of the first singular loci, a list of the NGFPs, including their position and stability coefficients, is given in Tab.\ \ref{tab.1}. 
For the linear split ($\alpha =0$) all regulators give rise to a NGFP with located at $g > 0,\lambda > 0$. The complex stability coefficients with positive real part indicate that this fixed point is a spiraling attractor which captures the RG flow in its vicinity as $k \rightarrow \infty$. In addition the Type II regulator may give rise to a second NGFP. This fixed point is a saddle point possessing one attractive and one repulsive eigendirection. Notably, the cases which possess this second fixed point coincide with the ones where $\eta^{\rm sing}$ does not screen the singular line $\lambda_1^{\rm sing}$, cf.\ Fig.\ \ref{singloci}.

The exponential split ($\alpha = 1$) gives rise to a similar picture. In this case the specific structure of the off-diagonal contributions implies that the Type II$_0$ and Type II$_2$ regulators actually coincide, giving rise to the same beta functions. Also this case gives rise to a NGFP with complex critical exponents which acts as a UV-attractor of the flow.  
The distinct critical exponents accompanying the spiraling NGFPs seen in the linear and exponential setting strongly support that they correspond to two distinct universality classes.\footnote{This agrees with the conclusion reached in \cite{Nink:2015lmq} where it was shown that the linear and exponential parametrization lead to Liouville theories with different central charge.} The Type II$_1$ case does not support a NGFP in the physically interesting region. This indicates that the system acts rather sensitive to a change of the eigenvalue spectrum in the scalar $\Nh$-$h$ sector. Thus we will limit our further considerations to the case $q_{\rm non-diag} = 0$. 

\begin{figure}[t!]
	\begin{center}
		\includegraphics[width=0.3\textwidth]{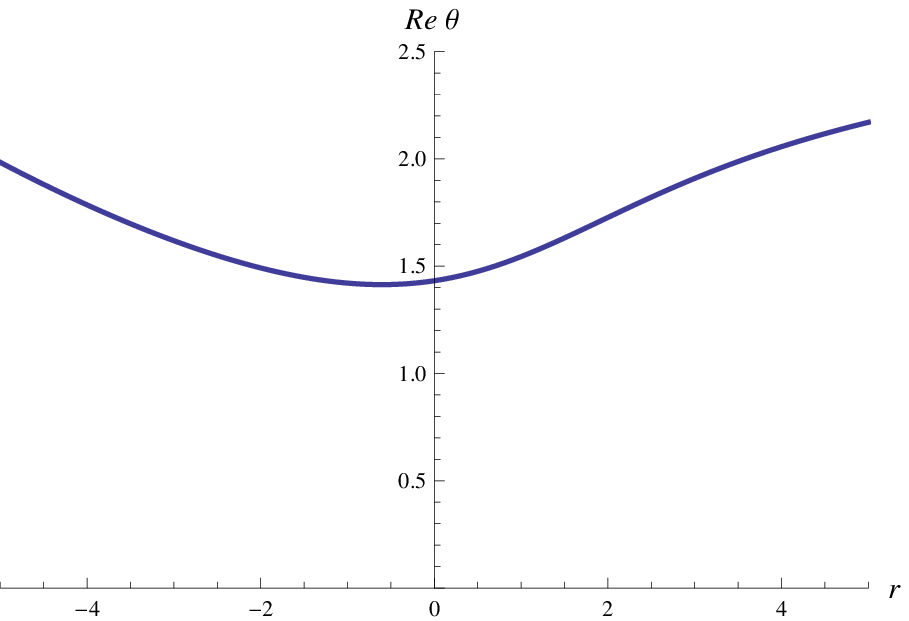} \;
		\includegraphics[width=0.3\textwidth]{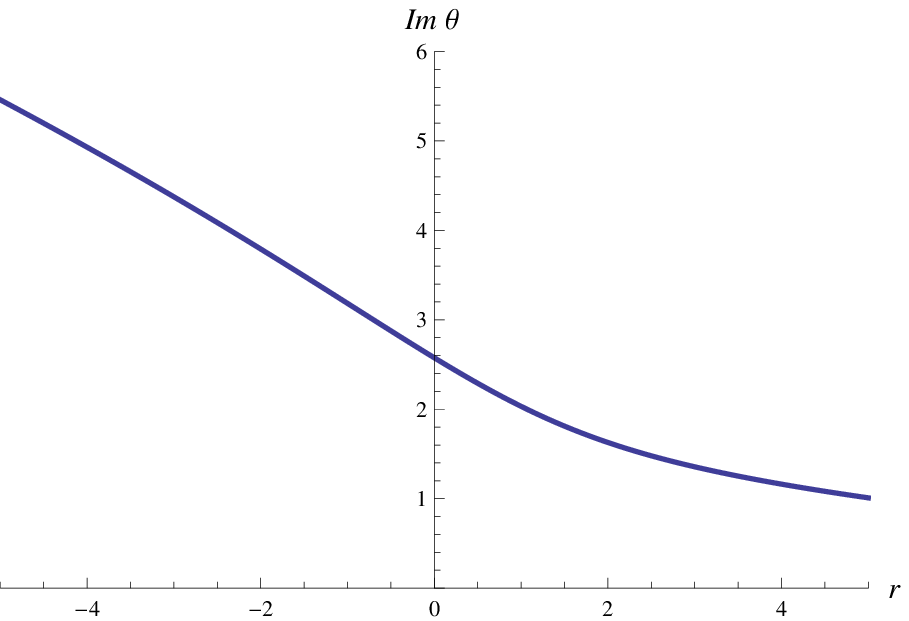} \;
		\includegraphics[width=0.3\textwidth]{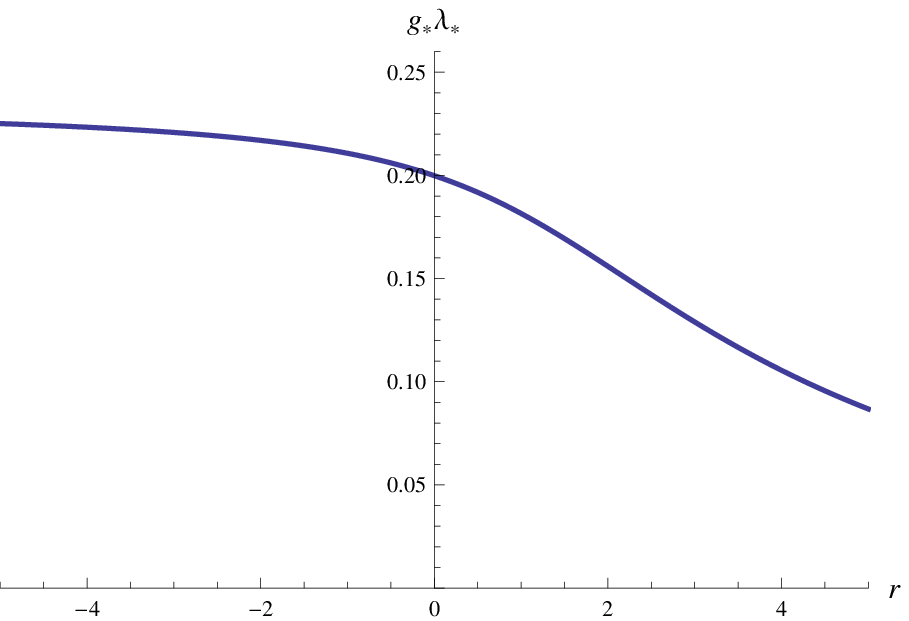} 
		\caption{\label{Fig.opt} Illustration of the $r$-dependence of the stability coefficients and universal product $g_* \lambda_*$ of the NGFP found for the linear split ($\alpha = 0$). The real part of $\theta$ possesses a minimum for $r_{\rm opt}^{\rm linear} = -0.605$ while Im$\theta$ and $g_* \lambda_*$ decrease monotonically for increasing $r$.}
	\end{center}	
\end{figure}
At this stage, it is natural to ask if there is a preferred value for the parameter $r$. Since a change in the regularization procedure should not  affect physical quantities, a natural selection criterion for $r$ is to minimize the sensitivity of these quantities with respect to this parameter. This optimization procedure \cite{Liao:1999sh,Litim:2000ci,Canet:2002gs,Pawlowski:2005xe}, may then be used to find a ``best value'' for the parameter $r$. 
Within the present computation natural candidates for investigating the $r$-dependence are the stability coefficients and the universal product $g_* \lambda_*$.
For a linear split their $r$-dependence is displayed in Fig.\ \ref{Fig.opt}. While Im$\theta$ and $g_*\lambda_*$ are monotonically decreasing as $r$ increases, Re$\theta$ develops a minimum located at $r_{\rm opt}^{\rm linear} = -0.605$. The corresponding values for the position and stability coefficients of the NGFP are listed in Tab.\ \ref{tab.1}. A comparison among the characteristic properties of the NGFP reveals that the ``optimized values'' turn out to be very close to the fixed point seen in the Type I regularization scheme. For the exponential split all physical quantities are monotonic functions of $r$. Thus in this case the principle of minimal sensitivity does not identify a preferred value for $r$.

\begin{figure}[t!]
	\begin{center}
		\includegraphics[width=0.48\textwidth]{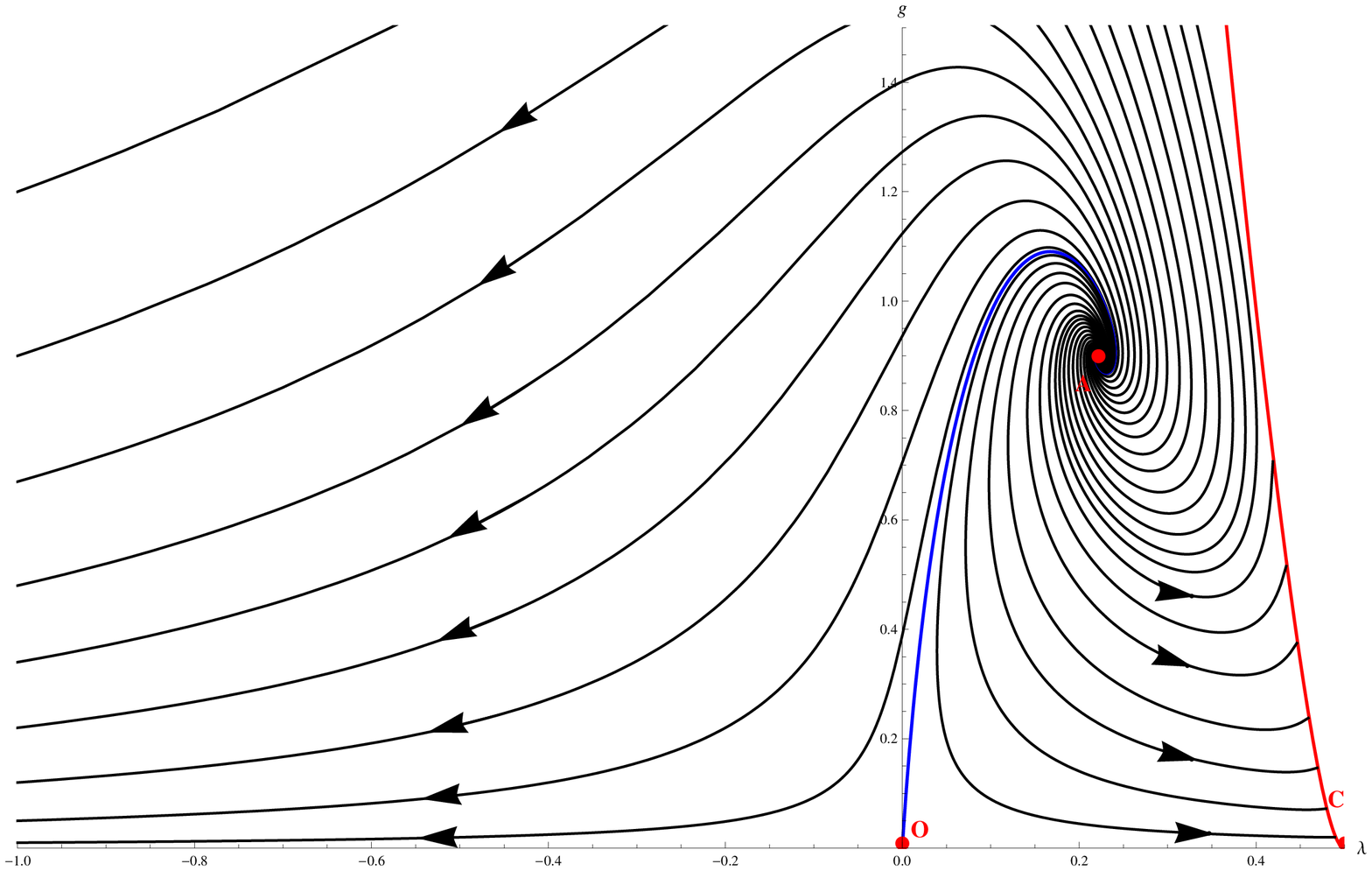} \;
		\includegraphics[width=0.48\textwidth]{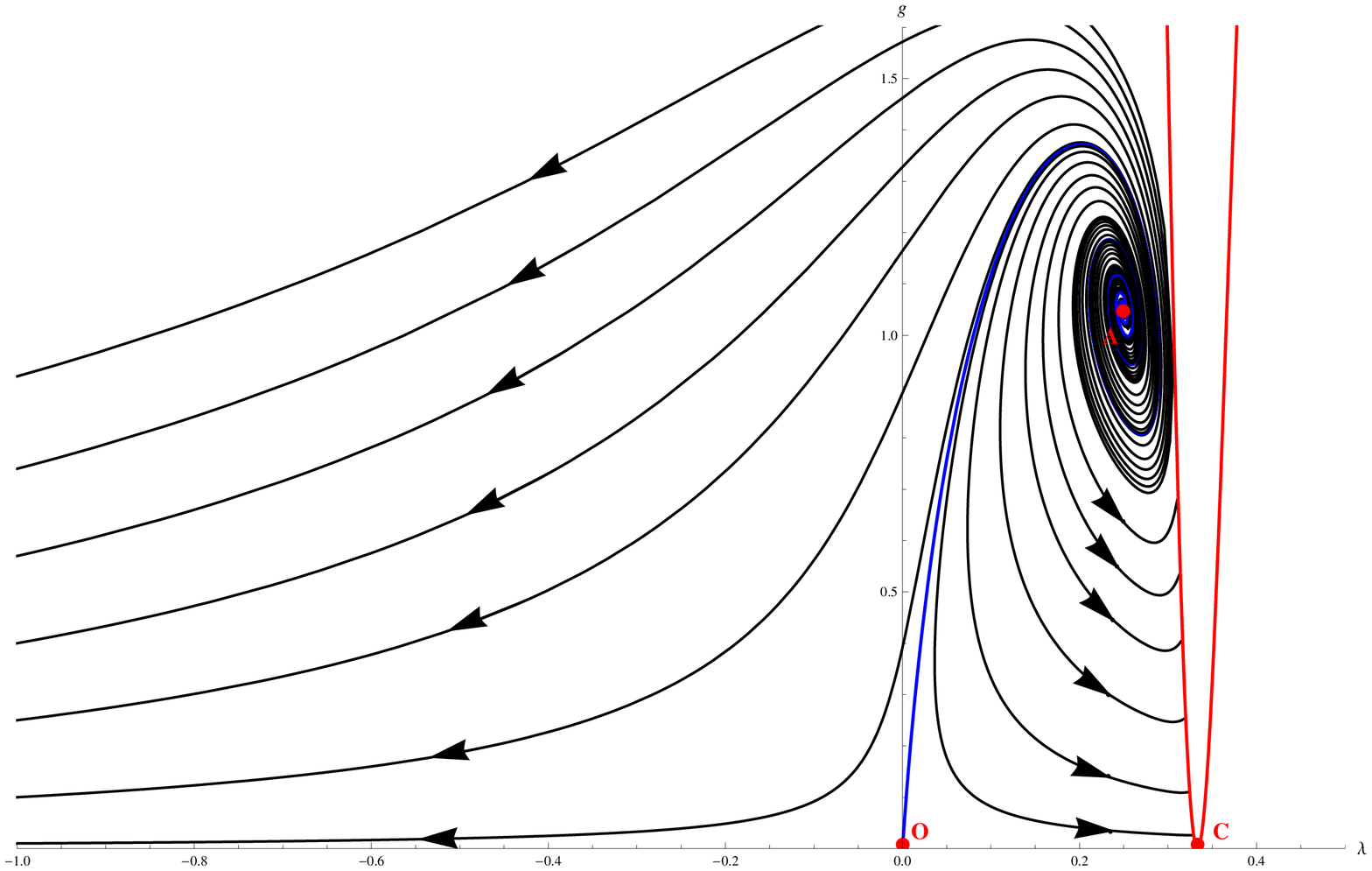} \\[3ex]
		\includegraphics[width=0.48\textwidth]{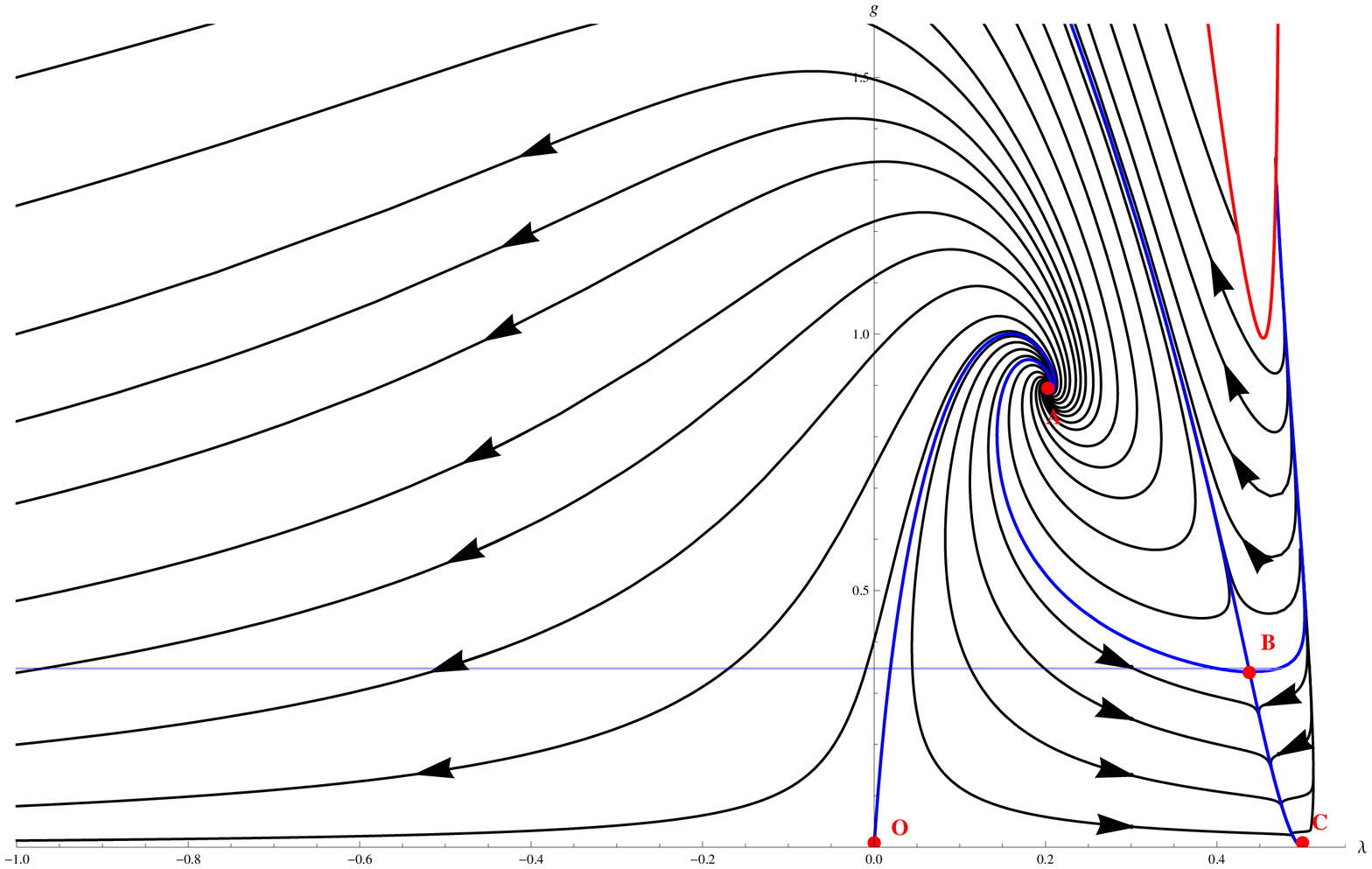} \;
		\includegraphics[width=0.48\textwidth]{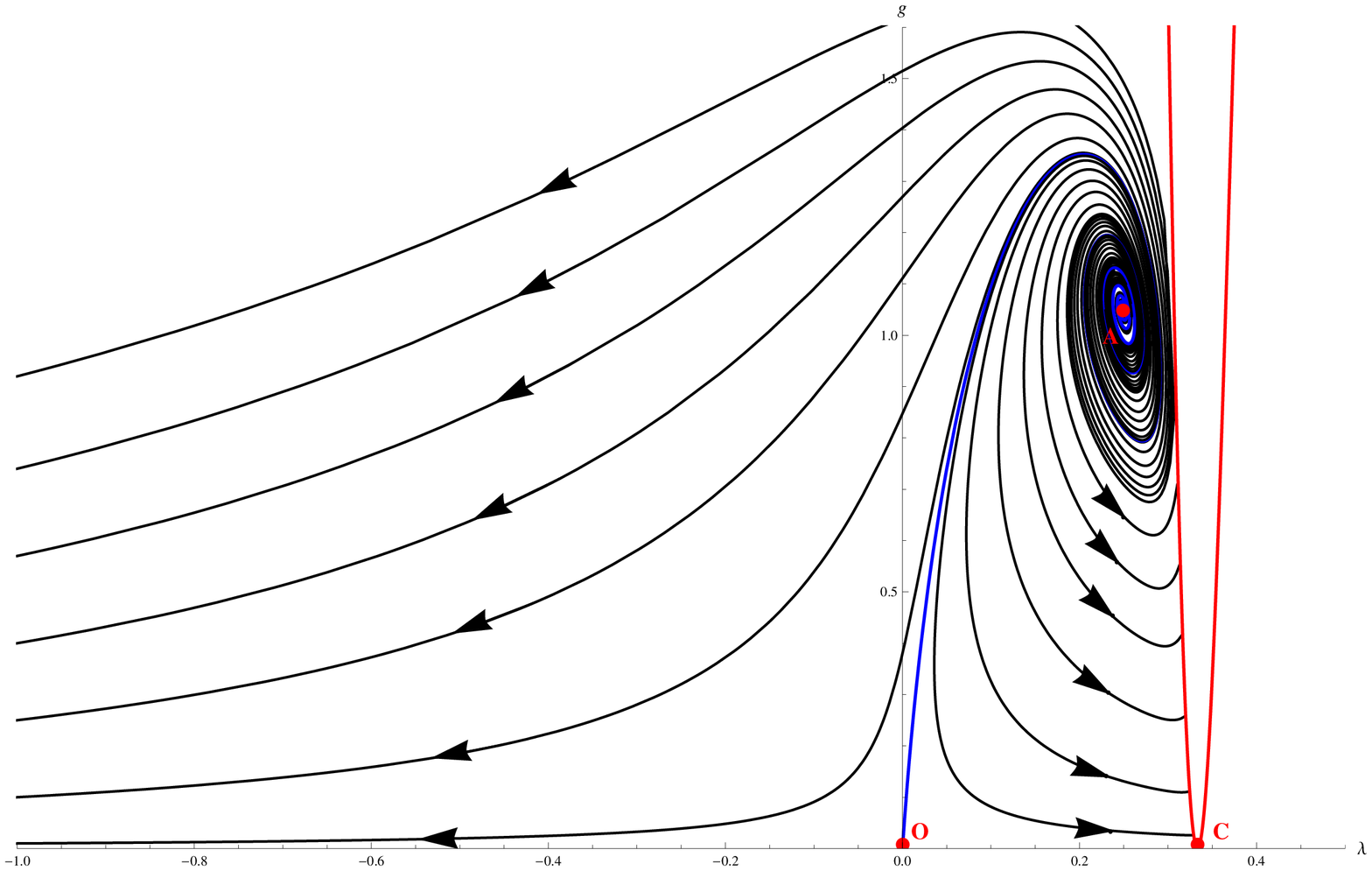}
		\caption{Phase diagrams obtained from integrating the beta functions \eqref{betafcts} for a Type I regulator ($r=0$, top line) and a Type II$_0$ regulator ($r=1$, bottom line). The first column gives the result for a linear split of the spatial metric ($\alpha = 0$) while the second column corresponds to an exponential split ($\alpha = 1$). All flows possess a NGFP providing the UV-completion of the RG trajectories.   \label{Fig.flow} }
	\end{center}	
\end{figure}
Based on the rather detailed discussion of their fixed points and singularity structure, it is rather straightforward to obtain the phase diagrams resulting from integrating the flow equations \eqref{betafcts} numerically. Our focus is on the physically interesting region where Newton's coupling is positive and containing the GFP. The resulting flows are shown in Fig.\ \ref{Fig.flow}. The left and right column display the results obtained with a linear ($\alpha = 0$) and exponential split ($\alpha = 1$), respectively. The top row stems from a Type I regulator while the bottom row uses the Type II$_0$ regularization scheme. In all cases the GFP is marked with $O$ while the non-Gaussian UV attractor carries the label $A$. The red lines mark the singular loci $\eta^{\rm sing}$, eq.\ \eqref{etasing}, while the blue lines connect the fixed points. All arrows point towards lower RG scales, i.e.\ in the direction of integrating out modes. 

As expected from the results given in Tab.\ \ref{tab.1}, the flow shows qualitative differences depending on whether the fixed point structure also contains the saddle point $B$ (Type II$_0$, linear) or just the NGFP $A$ (Type I, linear; exponential split). In the latter case, the phase diagram is dominated by the interplay of the GFP and NGFP. The NGFP controls the high-energy ($k \rightarrow \infty$) limit of all trajectories. Lowering the RG scale, the trajectories undergo a crossover to the GFP. In the vicinity of the GFP the trajectories develop a ``classical regime'' where the dimensionful couplings are almost independent of the RG scale $k$. Following \cite{Reuter:2001ag}, the solutions are classified  according to the value of the cosmological constant in this regime: trajectories located to the left and the right of the blue separatrix, give rise to a negative and positive value and are termed Type Ia and Type IIIa, respectively. The blue line separating these phases has a vanishing infrared value of $\Lambda_k$. The trajectories with $\Lambda_0 \le 0$ are complete in the sense that they are well-defined on the entire interval $k \in [0,\infty]$. The trajectories with a positive cosmological constant terminate at $\eta^{\rm sing}$ at a finite value $k$. The presence of the saddle point $B$ modifies this very last stage of the RG evolution. In this case the singularity $\eta^{\rm sing}$ is lifted and replaced by a RG trajectory connecting the fixed point $B$ and the quasi-fixed point $C$ located at $(\lambda,g) = (1/2,0)$. The flow then follows this line and is dynamically driven into the quasi-fixed point which provides the IR-completion of these trajectories.

\begin{figure}[t!]
	\begin{center}
		\includegraphics[width=0.48\textwidth]{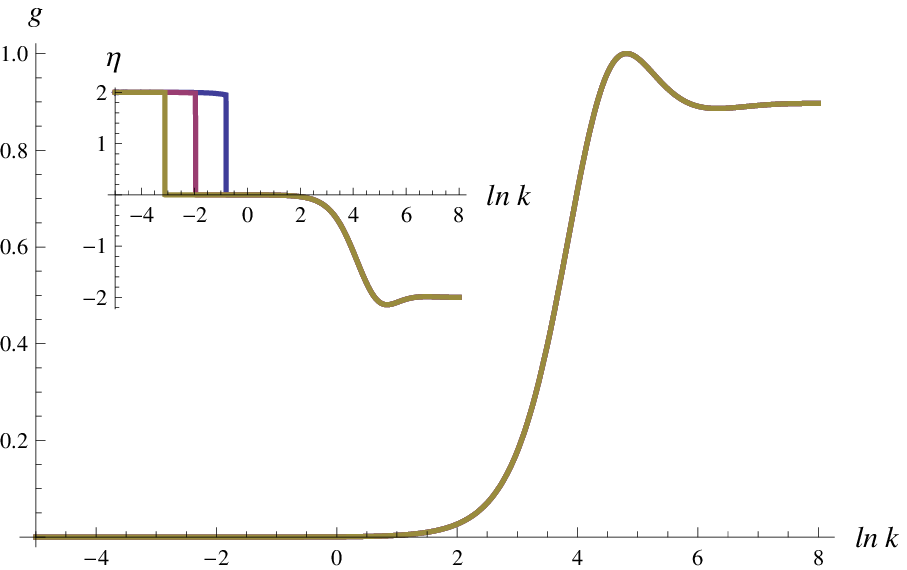} \;
		\includegraphics[width=0.48\textwidth]{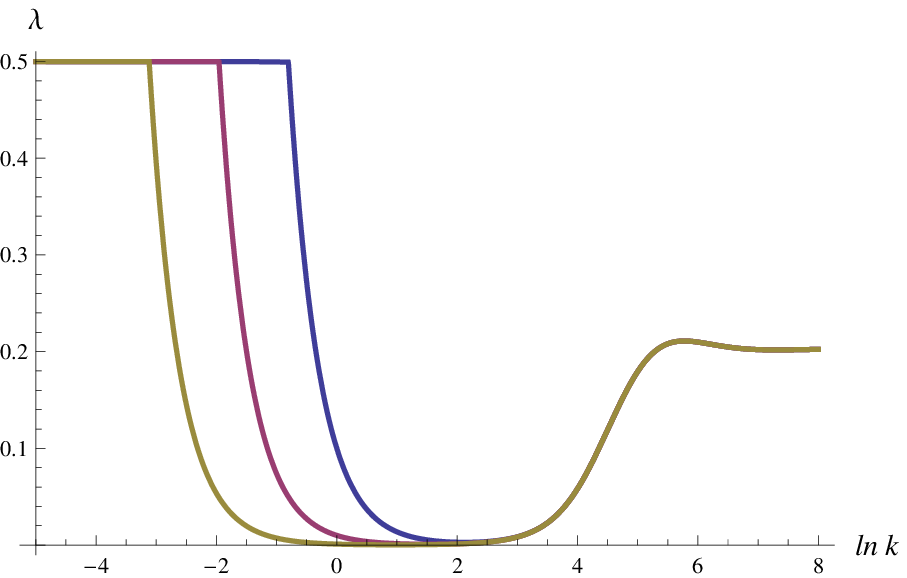} \\[3ex]
		\includegraphics[width=0.48\textwidth]{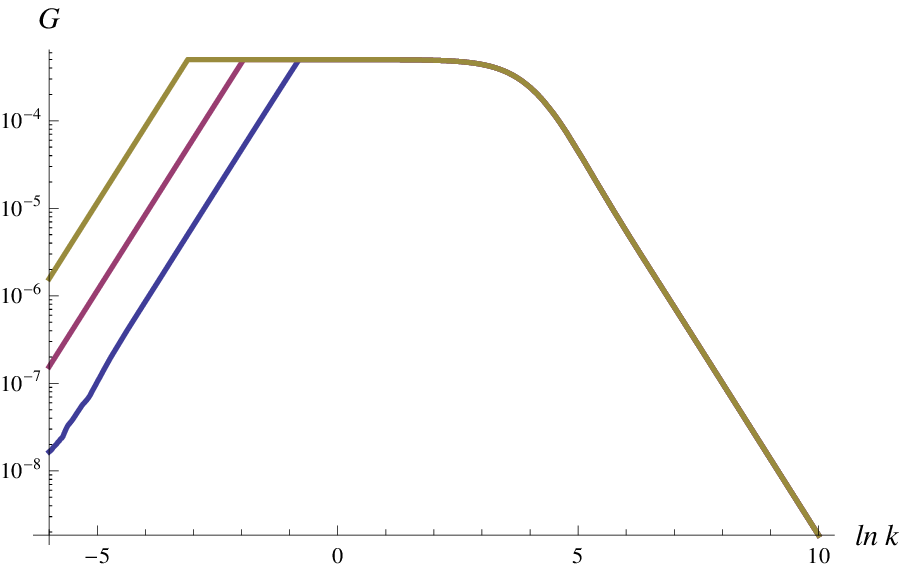} \;
		\includegraphics[width=0.48\textwidth]{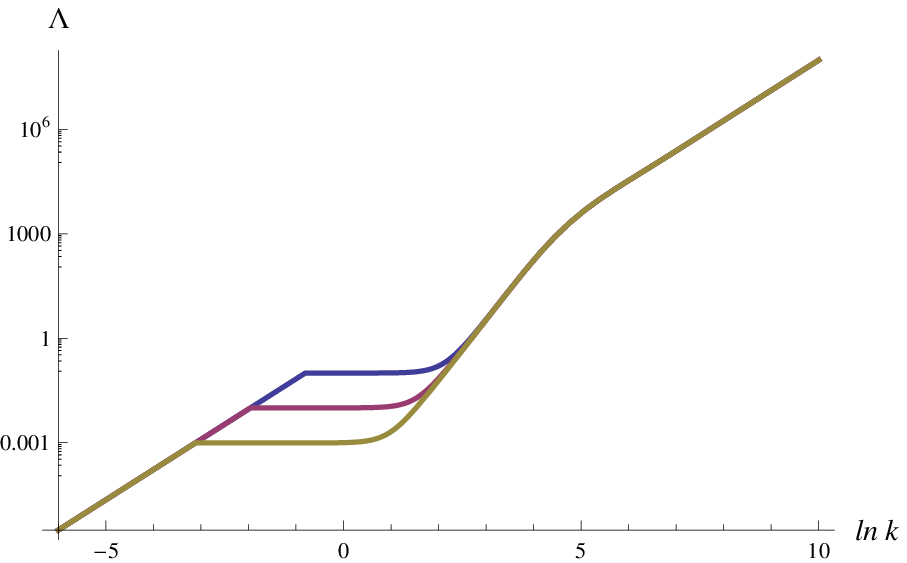}
		\caption{ \label{Fig.instability} Sample RG trajectories obtained from solving the flow equations for $\alpha = 0$ and a Type II$_0$ regulator for the initial conditions $g_{\rm init} = 0.0005$ and $\lambda_{\rm init} = 10^{-1}$ (blue line), $\lambda_{\rm init} = 10^{-2}$ (magenta line), and $\lambda_{\rm init} = 10^{-3}$ (gold line). For $k \rightarrow \infty$ the trajectories are governed by the NGFP. At intermediate scales one obtains a classical regime where the dimensionful $G_k$ and $\Lambda_k$ are independent of $k$. Once $\lambda \lesssim 0.5$ the gravitational instability sets in and drives the dimensionful Newton's coupling and cosmological constant to zero dynamically.} 
	\end{center}	
\end{figure}
Fig.\ \ref{Fig.instability} displays a set of sample trajectories obtained from integrating the beta functions for the linear split with a Type II$_0$ regularization scheme. The top line gives the scale-dependence of $g_k$ and $\lambda_k$. They interpolate between the NGFP for $\ln k \gtrsim 6$ and the quasi-fixed point $C$ for $\ln k \lesssim 1$. The lower diagrams recasts this flow in terms of the dimensionful couplings $G_k$ and $\Lambda_k$. In this way the three scaling regimes exhibited by the solutions become even more pronounced: in the NGFP regime one has $G_k \propto k^{-2}$ while $\Lambda_k \propto k^2$. The classical regime is situated around $\ln k \approx 0$. Once $\lambda_k$ reaches $\lambda_k \approx 1/2$ the flow enters into a new phase where $G_k$ and $\Lambda_k$ are driven to zero dynamically. The RG trajectories obtained from the other cases shown in Fig.\ \ref{Fig.flow} are similar. The only difference consists in the absence of the final low-energy phase. Here the flow terminates in the classical regime at a finite value of $k$.

At this stage, a cautious remark is in order. From the inset in the top-left diagram of Fig.\ \ref{Fig.instability} one finds that the new IR phase comes with an anomalous dimension of Newton's coupling $\eta_k \approx 2$. This has profound consequences for the regularization procedure. The regulator obtained from the prescription \eqref{cutoff} has the structure 
\be
\cR_k \propto (G_k)^{-1} \, k^2 \,  R^{(0)}(\Delta/k^2)
\ee
 where the profile function $R^{(0)}(\Delta/k^2)$ satisfies  $\lim_{k \rightarrow 0} R^{(0)}(\Delta/k^2) = 1$. In general, this asymptotic behavior ensures that the cutoff vanishes as $k\rightarrow 0$. The last property fails, however, if $G_k \propto k^2$ or, equivalently $\eta = 2$, as $k \rightarrow 0$. In this case the $k$-dependence of Newton's coupling cancels the $k^2$ term and the cutoff $\cR_k$ remains finite as $k \rightarrow 0$. As a result, a flow approaching the quasi fixed point $C$ may not integrate out all fluctuation modes, even though the limit $k \rightarrow 0$ is well-defined. Of course, all other phases displayed in Fig.\ \ref{Fig.instability} are unaffected by this peculiarity.

\section{Discussion and Outlook}
\label{sect.5}
In this work we have studied the gravitational renormalization group (RG) flow in the Arnowitt-Deser-Misner (ADM) formalism, utilizing backgrounds with a topology $S^1 \times S^d$. This investigation is mainly motivated through the Causal Dynamical Triangulation (CDT) program where this particular topology has been used extensively in order to study properties of the gravitational partition sum through Monte Carlo simulations. The detailed results reported in the main part of the manuscript and collected in the appendices provide an important stepping stone for comparing properties of the quantum spacetimes arising within the RG and CDT framework.

Our analysis focused on the scale-dependence of the (background) Newton's coupling and cosmological constant obtained from a projection of the functional renormalization group equation \eqref{FRGE}. The flow possesses a non-Gaussian fixed point (NGFP) suitable for rendering the theory asymptotically safe. The existence of this fixed point is robust with respect to changing the parametrization of the gravitational fluctuations and regularization procedure. The results summarized in Tab.\ \ref{tab.1} suggest that the NGFPs obtained from a linear and exponential split of the metric fluctuations belong to two different universality classes. The phase diagrams collected in Fig.\ \ref{Fig.flow} show that the flow emanating from the NGFP is connected to a classical regime where Newton's coupling and the cosmological constant are essentially independent of the renormalization group scale. For specific choices of the regularization schemes, the NGFP is supplemented by a second fixed point solution constituting a saddle point in the $g$-$\lambda$--plane. The interplay of the two fixed points alters the singularity structure of the beta functions which has profound consequences for the infrared behavior of the flow, see Fig.\ \ref{Fig.instability}.

The results obtained in this work are complementary to the ones reported in \cite{Biemans:2016rvp,Biemans:2017zca} which use a very similar construction on a background topology $S^1 \times T^d$. At the geometrical level, the key difference in these two classes of backgrounds is that $S^1 \times S^d$ possesses a Killing vector in the (Euclidean) time-direction which permits a Wick rotation to Lorentzian time without obtaining complex background geometries. At the level of the flow equation, the two projection schemes construct the flow of Newton's coupling based on two different interaction monomials: the $S^1 \times T^d$ background uses the kinetic terms for the gravitational fluctuations while $S^1 \times S^d$ uses a potential term build from the intrinsic curvature. At the classical Einstein-Hilbert action the relative coefficients of these terms are fixed by the diffeomorphism invariance. At the quantum level it is highly encouraging that the fixed point structure and phase diagrams resulting from these two projection schemes are almost identical.  

At this stage constructing gravitational RG flows within the ADM formalism has achieved a similar robustness as the one encountered in comparable computations using a covariant parametrization of the metric fluctuations \cite{Lauscher:2001ya,Reuter:2001ag,Litim:2003vp,Donkin:2012ud,Nagy:2013hka,Gies:2015tca}, at least at the background level. From the conceptual point of view, it is clear that the background field formalism based on the ADM decomposition gives rise a natural parametrization of the metric fluctuations. This parametrization is related to the one used in the covariant linear split in a non-linear way \cite{Biemans:2017zca}. In particular, combining the ADM-split and the exponential parametrization of the fluctuations on the spatial slices ensures that the signatures of the time part and spatial part of the physical metric are independently conserved when quantum fluctuations are taken into account. In this sense, the ADM formalism provides a more refined version of the exponential parametrization recently investigated in \cite{Ohta:2016jvw,Ohta:2016npm}. 
An interesting consequence associated with the different parametrization schemes for the metric fluctuations is that it may shift contributions of the RG flow from background vertices to vertices containing fluctuation fields. It would be very interesting to investigate this effect in approximations of the flow equation which also takes vertices containing the fluctuation fields into account.  We hope to come back to this point in the future.

The beta functions \eqref{betaexp} possess a ``gravitational instability'' associated with the singular line \eqref{fixedsing}. In Ref.\ \cite{Wetterich:2017ixo}, it has been suggested that this type of instability could provide a dynamical solution for the cosmological constant problem through strong RG effects in the infrared. In Fig.\ \ref{Fig.instability} we demonstrated that this mechanism may also work in the context of pure gravity. Our general analysis identified two possible caveats to this scenario. First, the gravitational instability may be shielded by a diverging anomalous dimension. This scenario is realized by the flows displayed in the top-left, top-right, and bottom-right diagram of Fig.\ \ref{Fig.flow}. Second, the anomalous dimension may acquire a value for which the implementation of the Wilsonian RG procedure requires a modification of the (standard) regularization scheme. It would be very interesting to see if dilaton gravity, where Newton's coupling is generated dynamically, resolves these difficulties in a natural way.


\bigskip
\noindent
{\bf Acknowledgements} \\
We thank J.\ Ambj{\o}rn, J.\ Biemans, A.\ Bonanno, A.\ Platania, R.\ Loll, R.\ Percacci and M.\ Reuter for helpful discussions. A.~K.\ acknowledges financial support from an Erasmus+ fellowship. The research of F.~S.
is supported by the Netherlands Organisation for Scientific
Research (NWO) within the Foundation for Fundamental Research on Matter (FOM) grants 13PR3137 and 13VP12.

\begin{appendix}
\section{The background geometry $S^1 \times S^d$}
\setcounter{equation}{0}
\label{App.A}
Throughout this work we simplify the computation by choosing a background geometry with topology $S^1 \times S^d$. The size of $S^1$ is kept fixed while the $S^d$ factor denotes a one-parameter family of spheres with arbitrary but fixed radius. Keeping the radius of $S^d$ as a free parameter allows to disentangle the volume term from an interaction term build from the intrinsic curvature scalar. Thus the background is sufficiently complex for distinguishing the two interaction monomials appearing on the left-hand-side of the projected flow equation.

In terms of the $D=d+1$-dimensional geometry, the background metric is given by $\gb_{\mu\nu} = {\rm diag}[1, \sib_{ij}]$, where $\sib_{ij}$ denotes the metric on the $d$-sphere. Since $S^d$ is a maximally symmetric space, the Riemann tensor $\Rb_{ijkl}$ and the Ricci tensor $\Rb_{ij}$
constructed from $\sib_{ij}$ can be related to the Ricci scalar $\Rb$ by
\be\label{maxsym}
\begin{split}
	\Rb_{ijkl} = & \, \frac{\Rb}{d(d-1)} \, \left( \sib_{ik} \,  \sib_{jl} - \sib_{il} \, \sib_{jk} \right) \, , \qquad
	\Rb_{ij} = \frac{1}{d} \, \sib_{ij} \, \Rb \, .
\end{split}
\ee
In addition the Ricci scalar is covariantly constant, $\Db_i \Rb = 0$.

A comparison with the decomposition \eqref{metcomp} then indicates that the values of the background ADM fields are given by
\be\label{backadm}
\Nb(\tau,y) = 1 \, , \qquad \Nb_i(\tau,y) = 0 \, , \qquad \sib_{ij}(\tau,y) = \sib_{ij}(y) \, .
\ee
The product nature of the background entails that $\sib_{ij}$ is independent of $\tau$ so that derivatives with respect to $\tau$ can be commuted freely with $\sib_{ij}$. Moreover, the Laplacians of the background spacetime,  $\Delta \equiv - \gb^{\mu\nu} \Db_\mu \Db_\nu$, and on the spatial slices, $ - \sib^{ij} \Db_i \Db_j$, are related by
\be\label{laplacians}
\Delta = - \p_\tau^2 - \sib^{ij} \Db_i \Db_j \, .
\ee
This identity allows to express the differential operators arising in the second variation of $\Gamma_k$ to differential operators constructed from the spacetime metric.

Upon implementing a suitable gauge fixing, all differential operators entering in the right-hand-side of the flow equation can be combined into $\Delta$. The resulting traces can then be evaluated using standard heat-kernel techniques \cite{Barvinsky:1985an,Vassilevich:2003xt}. For the present calculation it suffices to keep track of terms containing at most two covariant derivatives. For a coarse graining operator of the form
\be
\Box = \Delta + q \, \Rb \, , 
\ee
with $q$ being a numerical coefficient, the relevant terms in the expansion are given by
\be\label{heatmaster}
{\rm Tr}_i \, e^{-s \Box} = \frac{1}{(4 \pi s)^{D/2}} \int d\tau d^dy \Nb \sqrt{\sib} \left[ a_0 + a_2 \, s \, \Rb + \ldots \right] \, .
\ee
The terms indicated by the dots do not contributing to the present approximation. The heat-kernel coefficients $a_i$ depend on the index structure of the field and have been given, e.g., in \cite{Lauscher:2001ya}. The result is tabulated in Tab.\ \ref{Tab.2}. The terms proportional to $\delta_{2,d}$ arise from eigenmodes of the Laplacian which do not contribute to the fields appearing on the left-hand-side of the decompositions \eqref{dec1} and \eqref{dec2}.

\begin{table}[t!]
	\renewcommand{\arraystretch}{1.4}
	\begin{center}
		\begin{tabular}{|c||c|c|c|c|c|}
			\hline
			 & $S$ & $V$ & $T$ & $TV$ &   $TTT$  \\ \hline \hline
		$a_0$ & \hspace*{2mm} $1$ \hspace*{2mm} & \hspace*{2mm} $d$ \hspace*{2mm} & \hspace*{2mm} $\tfrac{1}{2} d(d+1)$ \hspace*{2mm} & \hspace*{2mm} $d-1$ \hspace*{2mm} & \hspace*{2mm} $\tfrac{1}{2} (d+1)(d-2)$ \hspace*{2mm} \\ \hline
		\multirow{2}{*}{$a_2$} &
		\multirow{2}{*}{$\tfrac{1-6q}{6}$} &
		\multirow{2}{*}{$\tfrac{(1-6q)d}{6}$} &
		\multirow{2}{*}{$\tfrac{(1-6q) d (d+1)}{12}$}  &
		$\tfrac{(d+2)(d-3)+6 \delta_{d,2}}{6d}$ &
		$\tfrac{(d+1)(d+2)(d-5+3\delta_{d,2})}{12(d-1)}$ \\
		& & & &
		$ - q(d-1)$ &
		$ - q \tfrac{(d-2)(d+1)}{2}$
		\\ \hline \hline
		\end{tabular}
		\caption{\label{Tab.2} Heat kernel coefficients for fields with differential constraints on the product manifold $S^1 \times S^d$. The labels indicate that the resulting $a_i$ has been computed for spacetime scalars ($S$), spatial vectors ($V$), spatial, symmetric tensors ($T$), transverse spatial vectors ($TV$), and symmetric, transverse-traceless spatial tensors ($TTT$), respectively.}
	\end{center}
\end{table}

\section{Evaluation of the second variations}
\setcounter{equation}{0}
\label{App.B}
This appendix summarizes the construction of the Hessian $\Gamma_k^{(2)}$ on a background $S^1 \times S^d$.
\subsection{The gravitational sector}
We start by expanding the gravitational action \eqref{EHaction} to second order in the fluctuation fields. In order to facilitate the computation, the Einstein-Hilbert action \eqref{EHaction} is decomposed into four interaction monomials 
\be\label{imon}
\renewcommand{\arraystretch}{1.2}
\begin{array}{ll}
	I_1 \equiv \int d\tau d^dy \, N \sqrt{\sigma} \, K_{ij} K^{ij} \, , \qquad &
	I_2 \equiv \int d\tau d^dy \, N \sqrt{\sigma} \, K^2 \, , \\
	I_3 \equiv \int d\tau d^dy \, N \sqrt{\sigma} \, R \, , \qquad &
	I_4 \equiv \int d\tau d^dy \, N \sqrt{\sigma} \, . \\
\end{array}
\ee
Moreover, we use the shorthand notations
\be
\int_y \equiv  \int d\tau \, d^dy \, \sqrt{\sib} \, , \qquad \mbox{and} \quad \Db^2 \equiv \sib^{ij} \Db_i \Db_j \, , 
\ee
to lighten our formulas. The expansion of the kinetic terms $I_1$ and $I_2$ around the background \eqref{background} yields the following terms quadratic in the fluctuation fields
\begin{eqnarray}
\delta^2 I_1 \! &=& \! \! \int_y  \left[\tfrac 12 ( \partial_{\tau} \hat{\sigma}_{ij} ) ( \partial_{\tau} \hat{\sigma}^{ij} ) - 2 (\bar{D}^i \hat{N}^j)(\partial_{\tau} \hat{\sigma}_{ij}) + (\bar{D}_i \hat{N}_j + \bar{D}_j \hat{N}_i) (\bar{D}^i \hat{N}^j) \right], \; \; \\
\delta^2 I_2 \! &=& \! \! \int_y  \left[\tfrac{1}{2} ( \partial_{\tau} \hat{\sigma} )^2 - 2 (\bar{D}^i \hat{N}_i)  (\partial_{\tau} \hat{\sigma}) + 2( \bar{D}^i \hat{N}_i )^2 \right] \, .
\end{eqnarray}
The result is independent of the decomposition of the spatial metric in terms of background and fluctuation fields \eqref{flucpara}: the parameter $\alpha$ encoding the difference between the linear and exponential split does not enter these expressions. The expansion of the potential terms $I_3$ and $I_4$ is given by
\begin{eqnarray}
\delta^2 I_3 \! \! &=& \! \! \int_y \, \Big[ \tfrac{1}{2} \hat{\sigma}_{ij} \bar{D}^2  \hat{\sigma}^{ij} - \hat{\sigma}^{ij} \bar{D}_j \bar{D}_k \hat{\sigma}^k_i + (2\hat{N} + \hat{\sigma}) \bar{D}_i \bar{D}_j \hat{\sigma}^{ij} - (2\hat{N} + \tfrac{1}{2}\hat{\sigma}) \bar{D}^2 \hat{\sigma} \qquad \quad \nonumber \\
&&\qquad
  + \bar{R} \left( \tfrac{d-2}{d} \hat{N}\hat{\sigma} + \tfrac{d^2-5d+8}{4d(d-1)} \hat{\sigma}^2 - \tfrac{d^2-3d+4}{2d(d-1)} \hat{\sigma}^{ij}\hat{\sigma}_{ij} + \alpha \, \tfrac{d-2}{4d} \,  \hat{\sigma}^{ij}\hat{\sigma}_{ij} \right)
\Big] \, , \\
\delta^2 I_4 \! \! &=& \! \! \int_y \,  \left[ \hat{N}\hat{\sigma} + \tfrac{1}{4} \hat{\sigma}^2 - \tfrac{1}{2} \left( 1 - \tfrac{\alpha}{2} \right) \hat{\sigma}^{ij}\hat{\sigma}_{ij} \right] \, .
\end{eqnarray}
The $\alpha$-dependence of $\delta^2 I_3 $ and $\delta^2 I_4$ indicates that this sector receives extra terms when the exponential split is implemented.

In order to facilitate the next steps of the computation, we perform a transverse-traceless decomposition of the fluctuation fields with respect to the background. For fluctuations in the shift vector $\Nh_i$ we use
\be\label{Tdec}
\Nh_i = u_i + \Db_i B \, , \qquad \Db^i u_i = 0 \, ,
\ee
where the constraint ensures that $u_i$ is transverse. Analogously, the fluctuations of the spatial metric are decomposed in a transverse-traceless tensor $h_{ij}$, a transverse vector $v_i$ and two scalars $\psi$ and $h$,
\be\label{TTdec}
\hat{\sigma}_{ij} = h_{ij} + \Db_i v_j + \Db_j v_i + \left(\Db_i \Db_j-\tfrac{1}{d}\, \bar{\sigma}_{ij} \, \Db^2\right)\psi + \tfrac{1}{d} \, \bar{\sigma}_{ij}h
\ee
satisfying 
\be
\Db^i h_{ij} = 0, \qquad \bar{\sigma}^{ij} h_{ij} = 0, \qquad \Db^i v_i = 0, \qquad \sib^{ij} \hat{\sigma}_{ij} = \sih = h \, .
\ee
This change of integration variables is accompanied by non-trivial Jacobians. Their form can be deduced by evaluating the scalar products
\be
\begin{split}
\int_y \hat{N}_i \hat{N}^i = & \int_y \Big[ u_i u^i - B\Db^2 B \Big] \, , \\
\int_y	\hat{\sigma}_{ij} \hat{\sigma}^{ij} = &  \int_y \Big[  h_{ij}h^{ij} - 2 v_i \left[ \Db^2 + \tfrac{\Rb}{d} \right] v^i + \tfrac{d-1}{d} \, \psi\left[\Db^2 \left(\Db^2+ \tfrac{\Rb}{(d-1)} \right)\right]\psi+\tfrac{1}{d}h^2 \Big] \, .
\end{split}
\ee
These Jacobians can then be taken into account by an additional field redefinition
\be\label{redef}
\begin{split}
B & \mapsto \big(-\Db^2\big)^{-1/2}B \, , \\
v_i & \mapsto \big(-\Db^2 - \tfrac{\Rb}{d}\big)^{-1/2} v_i \, , \\
\psi & \mapsto \big(-\Db^2\big)^{-1/2} \big(-\Db^2 - \tfrac{\Rb}{d-1}\big)^{-1/2} \psi \, .
\end{split}
\ee
This rescaling also ensures that all fields appearing in the decompositions have the same dimensionality. With a slight abuse in notation, we will work with the rescaled fields in the sequel.

Implementing the transverse-traceless decomposition in the kinetic terms and taking into account the rescaling \eqref{redef} yields
\begin{eqnarray}
\nonumber
	\delta^2 I_1 \! \! &=& \! \! \int_y \Big[ -\tfrac{1}{2} h_{ij} \, \partial_{\tau}^2 \, h^{ij} - v_i \, \partial_{\tau}^2 \, v^i + u_i \left[-\Db^2 -\tfrac{\Rb}{d}\right]u^i + 2B\left[-\Db^2-\tfrac{\Rb}{d}\right]B \qquad \\ \nonumber
	&& \qquad -  \tfrac{d-1}{2d} \psi \, \partial_{\tau}^2 \, \psi - \tfrac{1}{2d} \, h \, \partial_{\tau}^2 \, h - 2u_i \, \partial_{\tau}\left[-\Db^2 -\tfrac{R}{d}\right]^{1/2} v^i \\
\label{B.12}
	&& \qquad -
	 \tfrac{2(d-1)}{d} B\partial_{\tau}\left[-\Db^2-\tfrac{\Rb}{d-1}\right]^{1/2}\psi + \tfrac{2}{d} \, B \, \partial_{\tau} \left[-\Db^2\right]^{1/2} \, h \Big]  \, , \\
\label{B.13}	
\delta^2 I_2 \! \! &=& \! \! \int_y \Big[ -\tfrac{1}{2}h\partial_{\tau}^2 h - 2B\Db^2 B + 2B\, \partial_{\tau} \left[-\Db^2\right]^{1/2} \, h \Big] \, .
\end{eqnarray}
The potential terms become
\begin{eqnarray}
\nonumber
\delta^2 I_3 \! \! &=& \! \! \int_y \Big[ \tfrac{1}{2}h_{ij}\left[\Db^2 - \tfrac{d^2-3d+4}{d(d-1)} \Rb \right]h^{ij}
- \tfrac{d-2}{d} \, \Rb \, v_i v^i
+ \tfrac{(d-2)(d-1)}{2d^2} \psi\left[-\Db^2 - \Rb \right]\psi \qquad \\ \nonumber
&&   + \tfrac{(d-2)(d-1)}{2d^2} h\left[-\Db^2+\tfrac{d-4}{2(d-1)}\Rb\right]h
+ \tfrac{(d-2)(d-1)}{d^2} \, \psi\left[\Db^2(\Db^2+\tfrac{\Rb}{d-1})\right]^{1/2} h \\ \nonumber
&&  - \hat{N}\left[\tfrac{2d-2}{d}\Db^2 -\tfrac{d-2}{d} \Rb \right]h + \tfrac{2(d-1)}{d} \, \hat{N}\left[\Db^2(\Db^2 + \tfrac{\Rb}{d-1})\right]^{1/2} \psi \\ \label{B.14}
&& + \tfrac{d-2}{4d} \, \alpha \, \Rb \left(
h_{ij}h^{ij} + 2 \, v_i  v^i + \tfrac{d-1}{d} \, \psi^2 + \tfrac{1}{d} \, h^2
 \right)  \Big] \, ,  \\
 \label{B.15}
\delta^2 I_4 \! \! &=& \! \! \int_y \Big[ \hat{N}h + \tfrac{1}{4}h^2
- \tfrac{1}{2} \left(1 - \tfrac{\alpha}{2} \right) \left(
h_{ij}h^{ij} + 2 \, v_i  v^i + \tfrac{d-1}{d} \, \psi^2 + \tfrac{1}{d} \, h^2
\right) \Big] \, .
\end{eqnarray}

Based on these building blocks the part of $\Gamma_k^{\rm grav}$ quadratic in the fluctuation fields can be obtained as
\be
\delta^{2}\Gamma_k^{\rm grav} = \frac{1}{16 \pi G_k} \left( \delta^2 I_1 - \delta^2 I_2 - \delta^2 I_3 + 2 \Lambda_k \, \delta^2 I_4 \right) \, .
\ee
It is an intriguing observation that the kinetic terms do not depend on the specific decomposition of the spatial metric into background and fluctuation fields. The parameter $\alpha$ encoding the contributions from the exponential decomposition appears in the potential terms only.

\subsection{The gauge-fixing sector}
Inspecting eqs.\ \eqref{B.12} - \eqref{B.14} more closely, one recognizes that the spatial vector field $v^i$ has a rather peculiar dispersion relation which does not contain spatial derivatives. This feature persists if one adopts proper time gauge where the fluctuations in the lapse function and shift vector are zero. The gauge fixing \eqref{Sgf} cures this problem. This is demonstrated as follows.

Following the procedure of defining building blocks, it is natural to express the gauge-fixing action in terms of the two expressions
\be\label{I5I6}
\delta^2 I_5 = \int_x F^2,\qquad \delta^2 I_6 = \int_x F_i \, \bar{\sigma}^{ij} \, F_j.
\ee
The functionals $F$ and $F_i$ are linear in the fluctuation fields, so that these building blocks are quadratic in the fluctuations. This feature is indicated by labeling the blocks by $\delta^2 I_5$ and $\delta^2 I_6$.
Substituting the field decompositions \eqref{Tdec} and \eqref{TTdec} and invoking the rescaling \eqref{redef} the contributions \eqref{I5I6} are
\begin{eqnarray}
\delta^2 I_5 \! \! &=& \! \! \int_y \Big[\hat{N}\partial_{\tau}^2 h
+ B \,  \partial_{\tau}\left[ -\Db^2 \right]^{1/2} \, \big( h - 2 \Nh \big)
-\hat{N}\partial_{\tau}^2\hat{N}
 - B\Db^2 B - \tfrac{1}{4}h\partial_{\tau}^2h
  \Big] \, , \qquad \; \; \; \\ \nonumber
\delta^2 I_6 \! \! &=& \! \! \int_y \Big[
 v_i\left[-\Db^2-\tfrac{\Rb}{d}\right] v^i
 - u_i \partial_{\tau}^2 u^i
 +2 \, u_i \, \partial_{\tau}\left[-\Db^2 -\tfrac{\Rb}{d}\right]^{1/2} v^i
 - \tfrac{(d-2)^2}{4d^2} h \, \Db^2 \, h  \\ \nonumber
&& \; \;
- \tfrac{(d-1)^2}{d^2} \, \psi\left[\Db^2+\tfrac{\Rb}{d-1} \right]\psi
- B \, \partial_{\tau}^2 \, B
- \hat{N} \, \Db^2 \, \hat{N}
+ 2B \, \partial_{\tau} \left[- \Db^2 \right]^{1/2} \, \hat{N}  \\ \nonumber
&& \; \; + \tfrac{(d-2)(d-1)}{d^2} \psi \, \left[\Db^2 (\Db^2 + \tfrac{\Rb}{d-1})\right]^{1/2} \, h
 + \tfrac{2(d-1)}{d} \, B \, \partial_{\tau} \left[ -\Db^2-\tfrac{\Rb}{d-1} \right]^{1/2} \, \psi \qquad \qquad \\
&& \; \; + \tfrac{2(d-1)}{d}\psi\left[\Db^2 (\Db^2 + \tfrac{\Rb}{d-1})\right]^{1/2} \hat{N}
 -  \tfrac{d-2}{d} h \, \partial_{\tau} \left[ - \Db^2 \right]^{1/2} \, B
- \tfrac{d-2}{d} \,  h \, \Db^2 \hat{N}
 \Big]. 
\end{eqnarray}

The matrix elements appearing in the gravitational sector of the Hessian $\Gamma_k^{(2)}$ can then be read off from the combination
\be\label{gamma2grav}
\tfrac{1}{2} \delta^2 \Gamma_k^{\rm grav} + \Gamma_k^{\rm gauge-fixing} \, .
\ee
They are summarized in Tab.\ \ref{Tab.3}.
\begin{table}[t!]
		\renewcommand{\arraystretch}{1.7}
		\begin{center}
\begin{tabular}{|l|l|}
\hline \hline
 & $(32 \pi G_k)^{-1} \left[ \Gamma_k^{(2)} \right]_{ab}$ \\ \hline \hline
$h_{ij} h^{ij}$ & $\Delta  - 2 \left(1-\tfrac{\alpha}{2}\right) \Lambda_k + \left(\tfrac{d^2-3d+4}{d(d-1)} - \tfrac{d-2}{2d} \, \alpha \right) \Rb$
\\ \hline
$v_i \, v^i$ & $2 \left( \Delta - 2 \left(1-\tfrac{\alpha}{2}\right) \Lambda_k + \left(\tfrac{d-3}{d} - \tfrac{d-2}{2d} \, \alpha \right) \Rb \right)$
\\ \hline
$\psi^2$ & $\tfrac{d-1}{d} \left( \Delta - 2 \left(1-\tfrac{\alpha}{2}\right) \Lambda_k + \left( \tfrac{d-4}{d} - \tfrac{d-2}{2d} \alpha \right) \Rb \right)$
\\ \hline
$u_i \, u^i$ & $2 \left( \Delta - \tfrac{1}{d} \Rb \right)$
\\ \hline
$B^2$ & $2 \left( \Delta - \tfrac{2}{d} \Rb \right)$
\\ \hline
$h^2$ & $- \tfrac{d-2}{2d} \left( \Delta - 2 \Lambda_k + \left( \tfrac{d-4}{d} + \tfrac{\alpha}{d}  \right) \Rb \right) + \tfrac{\alpha}{d} \, \Lambda_k$
\\ \hline
$\Nh h$ & $-  \left( \Delta - 2 \Lambda_k +  \tfrac{d-2}{d} \Rb \right)$
\\ \hline
$\Nh^2$ & $2 \Delta$
\\ \hline \hline
$\bar{c}c$ & $(32 \pi G_k) \, \Delta$ \\ \hline
$\bar{b}^i b_i$ & $(32 \pi G_k) \, \left(\Delta - \tfrac{1}{d} \Rb\right)$
\\ \hline \hline
\end{tabular}
\caption{\label{Tab.3} List of non-zero matrix elements appearing in the Hessian $\Gamma_k^{(2)}$. The gravitational sector is obtained from the combination \eqref{gamma2grav} while the ghost contributions given in the last two lines arise from \eqref{ghosts}. Each off-diagonal entry is accompanied by a suitable entry with the order of the fields reversed.}
\end{center}
\end{table}
Owed to the specific choice of the gauge-fixing all differential operators combine into Laplacians of the $D=d+1$-dimensional background manifold \eqref{laplacians}. Based on the Hessian given in Tab.\ \ref{Tab.3}, the regulator $\cR_k$ is constructed from the prescription \eqref{cutoff}. This completes the explicit construction of the traces appearing on the right-hand-side of the flow equation.

\section{Evaluation of the operator traces}
\setcounter{equation}{0}
\label{App.C}
Based on the results collected in Tab.\ \ref{Tab.3}, the operator traces appearing on the right-hand-side of the flow equation can be evaluated by combining standard Mellin-transform techniques, reviewed in \cite{Codello:2008vh}, with the heat-kernel formula \eqref{heatmaster}. We start by giving a master formula for evaluating the operator traces in App.\ \ref{App.C1} while the explicit results are collected in App.\ \ref{App.C2}.
\subsection{The master formula}
\label{App.C1}
The contributions of the operator traces appearing in the FRGE are conveniently expressed in terms of the dimensionless threshold functions \cite{Reuter:1996cp}
\be\label{tresholdfcts}
\begin{split}
	\Phi^p_n(w) \equiv \frac{1}{\Gamma(n)} \int_0^\infty dz \, z^{n-1} \, \frac{R^{(0)}(z) - z R^{(0)\prime}(z)}{[z+ R^{(0)}(z) + w]^p} \, , \\
	\widetilde{\Phi}^p_n(w) \equiv \frac{1}{\Gamma(n)} \int_0^\infty dz \, z^{n-1} \, \frac{R^{(0)}(z)}{[z+ R^{(0)}(z) + w]^p} \, .
\end{split}
\ee
For a cutoff of Litim type \cite{Litim:2003vp}, $R_k=(k^2-\Delta_s)\,\theta(k^2-\Delta_s)$, to which we resort in the main part of the paper the integrals in the threshold functions can be evaluated analytically
\be\label{thresholdfcts}
\Phi^p_n(w) \equiv \frac{1}{\Gamma(n+1)} \, \frac{1}{(1+w)^p} \, , \qquad
\widetilde{\Phi}^p_n(w) \equiv \frac{1}{\Gamma(n+2)} \, \frac{1}{(1+w)^p} \, .
\ee

Apart from the $\Nh$-$h$ sector, all Hessians given in Tab.\ \ref{Tab.3} have the structure\footnote{Since constant prefactors multiplying the Hessians drop out of the trace, we do not include them in the definition of $\cP$.}
\be\label{propagator}
\cP = \left( 32 \pi G_k \right)^{-s} \left( \Box + w \Lambda_k + c \Rb \right) \mathbb{1} \, ,
\ee
where $\mathbb{1}$ is the unit on the internal space and the parameter $s = 0,1$ in the ghost sector and gravitational sector respectively. The contributions of the fluctuation fields to the flow can then be obtained from the following master formula
\begin{eqnarray}\nonumber
{\rm Tr} \left[ \cP^{-1} \p_t \cR_k \right] \! \! & = & \! \!
\frac{k^D}{(4\pi)^{D/2}} \int_y \Big[
a_0 \left(2 \, \Phi^1_{D/2}(\wtw) - s \, \eta \, \widetilde{\Phi}^1_{D/2}(\wtw) \right) \\ \nonumber
&& \qquad \quad
+ a_2 \left(2 \, \Phi^1_{D/2-1}(\wtw) - s \, \eta \, \widetilde{\Phi}^1_{D/2-1}(\wtw) \right) \tfrac{\Rb}{k^2}
\\ \label{mastertr}
&& \qquad \quad
- c \, a_0 \left(2 \, \Phi^2_{D/2}(\wtw)
- s \, \eta \, \widetilde{\Phi}^2_{D/2}(\wtw) \right) \tfrac{\Rb}{k^2}
\Big] \, .
\end{eqnarray}
The heat-kernel coefficients $a_n$ depend on the spin and endomorphism parameter $q$ and are tabulated in Tab.\ \ref{Tab.2}. The anomalous dimension $\eta$ is defined in \eqref{dimless} and $\wtw \equiv w \lambda_k$ is the dimensionless version of the combination $w \Lambda_k$. For a Type II regulator $c=0$ by construction.

\subsection{Explicit results for the operator traces}
\label{App.C2}
Based on the master formula \eqref{mastertr}, it is rather straightforward to find the contributions of the terms listed in Tab.\ \ref{Tab.3} to the projection of the RG flow.
\subsubsection{Diagonal terms in the gravitational sector}
The master formula \eqref{mastertr} indicates that in the gravitational sector the threshold functions always appear in the combination
\be\label{qfct}
q_n^p(\wtw) \equiv 2 \, \Phi^p_{n}(\wtw)
-  \eta \, \widetilde{\Phi}^p_{n}(\wtw) \, .
\ee
In the diagonal sector, the argument $\wtw$ captures the dependence of the traces on the cosmological constant and it is convenient to set
\be\label{cosmo1}
\wtw_d \equiv (\alpha - 2) \, \lambda \, . 
\ee
Notably, the value $\alpha = 2$ (which does not correspond to an exponential split) is special since the $\lambda$-dependence of the flow is restricted to the off-diagonal sector. Making use of the master formula \eqref{mastertr} the contributions of the component fields associated with the spatial metric are
\begin{align}
{\rm Tr}_{h_{ij}h^{ij}} &= \tfrac{k^D (d+1)(d-2)}{2(4\pi)^{D/2}}\int_x \Bigl[q^1_{D/2} + \left(\tfrac{d^2-3d+4}{d(d-1)}-\tfrac{d-2}{2d}\alpha\right) (r-1)  \, q^2_{D/2} \, \tfrac{\bar{R}}{k^2}
  \nonumber\\
& \qquad \qquad \qquad  +\left(\tfrac{(d+2)(d-5+3\delta_{d,2})}{6(d-1)(d-2)} - r\left(\tfrac{d^2-3d+4}{d(d-1)}-\tfrac{d-2}{2d}\alpha\right)\right) \, q^1_{D/2-1} \, \tfrac{\bar{R}}{k^2}  \Bigr] \, ,  \nonumber\\
{\rm Tr}_{v_{i}v^{i}} &= \tfrac{k^D (d-1)}{(4\pi)^{D/2}}\int_x \Bigl[q^1_{D/2}+ \left(\tfrac{d-3}{d}-\tfrac{d-2}{2d}\alpha\right) \, (r-1) \, q^2_{D/2} \, \tfrac{\bar{R}}{k^2} 
\nonumber\\
& \qquad \qquad \qquad +\left(\tfrac{(d+2)(d-3)+6\delta_{d,2}}{6d(d-1)} - r\left(\tfrac{d-3}{d}-\tfrac{d-2}{2d}\alpha\right)\right) \, q^1_{D/2-1} \, \tfrac{\bar{R}}{k^2} \Bigr] \, , \nonumber\\
{\rm Tr}_{\psi\psi} &= \tfrac{k^D }{(4\pi)^{D/2}}\int_x \Bigl[q^1_{D/2}+ \left(\tfrac{d-4}{d}-\tfrac{d-2}{2d}\alpha\right) \, (r-1) \, q^2_{D/2}
\, \tfrac{\bar{R}}{k^2} 
\nonumber\\ \label{metcont}
&\qquad\qquad \qquad +\left(\tfrac16 -r\left(\tfrac{d-4}{d}-\tfrac{d-2}{2d}\alpha\right) \right) \, q^1_{D/2-1} \, \tfrac{\bar{R}}{k^2} \Bigr] \, . 
\end{align}
Here the argument of all threshold functions is given by \eqref{cosmo1}.

The traces containing fields $B$ and $u_i$, arising from the decomposition of the shift vector, do not receive a contribution from the cosmological constant. In this case the threshold functions $q_n^p$ are evaluated at zero argument and the resulting trace contributions read
\begin{align}
{\rm Tr}_{BB} &= \tfrac{k^D }{(4\pi)^{D/2}}\int_x \Big[ q^1_{D/2} +\left(\tfrac16 + \tfrac{2r}d \right) \, q^1_{D/2-1} \, \tfrac{\bar{R}}{k^2} - \tfrac{2}{d} \, (r-1) \, q^2_{D/2} \, \tfrac{\bar{R}}{k^2} \Big] \, ,  \nonumber\\
\label{contshift}
{\rm Tr}_{u_{i}u^{i}} &= \tfrac{k^D (d-1)}{(4\pi)^{D/2}}\int_x \Bigl[q^1_{D/2} 
+\left(\tfrac{(d+2)(d-3)+6\delta_{d,2}}{6d(d-1)} + \tfrac{r}d\right) \, q^1_{D/2-1} \, \tfrac{\bar{R}}{k^2}
- \tfrac{1}{d} \, (r-1) \, q^2_{D/2} \, \tfrac{\bar{R}}{k^2} \Bigr] \, . 
\end{align}
The results \eqref{metcont} and \eqref{contshift} complete the evaluation of the operator traces appearing in the diagonal part of the gravitational sector.
\subsubsection{Off-diagonal terms in the gravitational sector}
The final contribution in the gravitational sector originates from the sector spanned by the fluctuations of the lapse function $\Nh$ and the conformal modes of the spatial metric $h$. In this sector the fluctuations span a full $2\times2$ block matrix in field space. Following the Type I regularization procedure, one can construct the cutoff $\cR_k$ in this sector and subsequently invert this block matrix in field space. The propagators appearing in this procedure are not of the simple form \eqref{propagator} but contain terms quadratic in the spacetime Laplacians. Using the specific properties of the Litim-type regulator \eqref{thresholdfcts}, the result can still be expressed in terms of the threshold functions. For the Type I regulator, the resulting contribution is given by
\begin{eqnarray}\nonumber
 {\rm Tr}_s \! \! & = & \! \!  \tfrac{k^D}{(4\pi)^{D/2}} \int_y \Big[ \left(2-\tfrac{3d-2+\alpha}{d-1} \lambda \right) \left(q^1_{D/2}(\wtw_s) + \tfrac{1}{6} \, q^1_{D/2-1}(\wtw_s) \, \tfrac{\Rb}{k^2} \right) \\ \label{scalartypeI}
  &&  - \tfrac{d-2}{2d(d-1)} \left( \tfrac{3 d^2 - 7 d + 4}{d-1} + \alpha - 8 d  \lambda + \tfrac{2d(3d+\alpha)}{d-1} \lambda^2 \right) \, q^2_{D/2}(\wtw_s) \, \tfrac{\Rb}{k^2}
 \Big] \, ,
\end{eqnarray}
where
\be
\wtw_s \equiv - \tfrac{3d-2+\alpha}{d-1} \, \lambda + \tfrac{2d}{d-1} \, \lambda^2 \, .
\ee

The implementation of a Type II regulator in this sector is non-trivial. This complication can be traced back to the feature that, in contrast to the other fields, the terms in the off-diagonal sector come with different relative coefficients between the Laplacians $\Delta$ and the intrinsic curvature $\Rb$.
Inspecting the structure of $\Gamma_k^{(2)} + \cR_k$ reveals that there are two natural candidates for choosing an endomorphism in the regulator
\be\label{scalarTypeII}
\begin{split}
\Box_1 = & \, \Delta + \tfrac{d-2}{d} \Rb \, , \qquad
\Box_2 = \Delta + \tfrac{d-4+\alpha}{d} \Rb \, .
\end{split}
\ee
The inclusion of the non-zero endomorphism slightly modifies the contribution of the scalar sector. For the choice $\Box_1$ the trace evaluates to
\begin{eqnarray}\nonumber
{\rm Tr}_s \! \! & = & \! \!  \tfrac{k^D}{(4\pi)^{D/2}} \int_y \Big[ \left(2-\tfrac{3d-2+\alpha}{d-1} \lambda \right) \left(q^1_{D/2}(\wtw_s) -\tfrac{5d-12}{6d} \, q^1_{D/2-1}(\wtw_s) \, \tfrac{\Rb}{k^2} \right) \\ \label{scalartypeIIa}
&&  + \tfrac{d-2}{2d} \left( \tfrac{d- \alpha}{d-1} - \tfrac{4  (d-2 + \alpha)}{d-1} \lambda + \tfrac{2 (2 d^2 - 8 d + 4  + (5d- 4) \alpha + \alpha^2)}{(d-1)^2} \lambda^2 \right) \, q^2_{D/2}(\wtw_s) \, \tfrac{\Rb}{k^2}
\Big] \, ,
\end{eqnarray}
while for the second choice, $\Box_2$ one obtains
\begin{eqnarray}\nonumber
{\rm Tr}_s \! \! & = & \! \!  \tfrac{k^D}{(4\pi)^{D/2}} \int_y \Big[ \left(2-\tfrac{3d-2+\alpha}{d-1} \lambda \right) \left(q^1_{D/2}(\wtw_s) -\tfrac{5d-24+6\alpha}{6d} \, q^1_{D/2-1}(\wtw_s) \, \tfrac{\Rb}{k^2} \right) \\ \nonumber
&&  + \tfrac{1}{d} \Big( \tfrac{d^2-10d+8 + (3d - 2) \alpha}{2(d-1)}
- 2 \, \tfrac{d^2-10d+8 + (4d - 6) \alpha + \alpha^2}{d-1} \lambda
\\ && \qquad \label{scalartypeIIb}
 + \tfrac{2 d^3 - 22 d^2 + 36d -16
 	 + (10 d^2  - 34 d + 20 ) \alpha + (7 d-8 ) \alpha^2 +
 	 \alpha^3}{(d-1)^2} \lambda^2 \Big) \, q^2_{D/2}(\wtw_s) \, \tfrac{\Rb}{k^2}
\Big] \, .
\end{eqnarray}
This result completes the evaluation of the operator traces appearing in the gravitational sector.
\subsubsection{Ghost contributions}
In the ghost sector all threshold functions are evaluated at zero argument and there are no terms containing $\eta$. The explicit contributions of the traces is given by
\begin{eqnarray}\label{trghost}
- {\rm Tr}_{\bar{c}c} \! \! & = & \! \! - \tfrac{k^D}{(4\pi)^{D/2}} \int_y \Big[ 2 \, \Phi^1_{D/2} + \tfrac{1}{3} \, \Phi^1_{D/2-1} \, \tfrac{\Rb}{k^2} \Big] \, , \\ \nonumber
- {\rm Tr}_{\bar{b}b} \! \! & = & \! \! - \tfrac{k^D}{(4\pi)^{D/2}} \int_y \Big[
2 d \, \Phi^1_{D/2}
+ \left(\tfrac{1}{3} d + 2 r \right) \, \Phi^1_{D/2-1} \, \tfrac{\Rb}{k^2}
+ 2 \left(1-r\right) \Phi^2_{D/2} \, \tfrac{\Rb}{k^2}
\Big] \, .
\end{eqnarray}
Together with the formulas \eqref{metcont}, \eqref{contshift}, and \eqref{scalartypeI}, this result completes the evaluation of all traces appearing on the right-hand-side of the FRGE. 
\end{appendix}

\end{document}